\documentclass[preprint,12pt]{elsarticle}
\graphicspath{../}
\usepackage[english]{babel}
\usepackage[utf8x]{inputenc}
\usepackage{listings}
\usepackage{color}
\usepackage{algorithm}
\usepackage{algorithmic}
\usepackage{verbatim}
\usepackage{soul} 
\usepackage[toc,page]{appendix}
\usepackage{amsfonts}
\usepackage[colorlinks = true,linkcolor =red, urlcolor = blue]{hyperref}
\usepackage{caption}
\usepackage{subcaption}
\usepackage{url}
\usepackage{hyperref}
\usepackage{makecell}
\usepackage{chngcntr}
\usepackage{apptools}
\usepackage{ntheorem}
\usepackage{multirow}
\usepackage{tabularx}
\usepackage{xcolor}
\definecolor{darkred}{RGB}{139, 0, 0}

\usepackage{float}
\usepackage{shuffle}
\usepackage{graphicx}
\usepackage{diagbox}

\AtAppendix{\renewtheorem{theorem}{Theorem A.}}

\newcommand{\tri}[1]{{\left\vert\kern-0.25ex\left\vert\kern-0.25ex\left\vert #1 
    \right\vert\kern-0.25ex\right\vert\kern-0.25ex\right\vert}}

\usepackage{eqparbox}

\usepackage{amsmath}
\usepackage[colorinlistoftodos]{todonotes}
\usepackage[colorlinks=true,
            linkcolor = red,
            urlcolor=blue]{hyperref}
\usepackage{listings}
\usepackage{url}
\usepackage{graphicx}
\usepackage{subcaption}
\usepackage{cleveref}

\usepackage{amssymb}

\usepackage{booktabs}

\journal{Pattern Recognition}

\newsavebox{\measurebox}

\begin{document}

\begin{frontmatter}



\title{NODE-ImgNet: a PDE-informed effective and robust model for image denoising}

\author[clyde]{Xinheng Xie}
\author[clyde]{Yue Wu}
\author[ucl,ATI]{Hao Ni}
\author[ok,uga,*]{Cuiyu He}

\affiliation[clyde]{organization={Department of Mathematics and Statistics, University of Strathclyde},
            addressline={26 Richmond St}, 
            city={Glasgow},
            postcode={G1 1XQ}, 
            country={UK}}
 \affiliation[ucl]{organization={Department of Mathematics, University College London},
             addressline={25 Gordon St},
             city={London},
             postcode={WC1H 0AY},
             country={UK}}

 \affiliation[ATI]{
             organization={The Alan Turing Institute},
             addressline={2QR, John Dodson House, 96 Euston Rd},
             city={London},
             postcode={NW1 2DB},
             country={UK}}            

\affiliation[ok]{organization={Department of Mathematics, Oklahoma State University},
            addressline={401}, 
            city={Stillwater, OK},
            postcode={74078}, 
            country={USA}}

\affiliation[uga]{organization={Department of Mathematics, University of Georgia},
            addressline={1023 D. W. Brooks Drive}, 
            city={Athens, GA},
            postcode={30605}, 
            country={USA}}
\affiliation[*]{Corresponding author: cuiyu.he@okstate.edu}
\begin{abstract}
Inspired by the traditional partial differential equation (PDE) approach for image denoising, we propose a novel neural network architecture, referred as NODE-ImgNet, that combines neural ordinary differential equations (NODEs) with convolutional neural network (CNN) blocks.
NODE-ImgNet is intrinsically a PDE model, where the dynamic system is learned implicitly without the explicit specification of the PDE. This naturally circumvents the typical issues associated with introducing artifacts during the learning process. By invoking such a NODE structure, which can also be viewed as a continuous variant of a residual network (ResNet) and inherits its advantage in image denoising, our model achieves enhanced accuracy and parameter efficiency.  In particular, our model exhibits consistent effectiveness in different scenarios, including denoising gray and color images perturbed by Gaussian noise, as well as real-noisy images, and demonstrates superiority in learning from small image datasets.

\end{abstract}


\end{frontmatter}


\section{Introduction}
Various factors can cause noise in images. The most common ones are sensor noise \cite{zhang2017improved}, compression noise \cite{nam2016holistic}, optical noise 
\cite{kuang2017single}, etc. 
Referring as the process of removing unwanted noise or artifacts from an image, image denoising is an important task in image processing and computer vision.  It is critical for improving the quality, accuracy, and usefulness of images in a wide range of applications.
Image denoising techniques have been extensively explored both in the industry and academia in the last several decades, see e.g., \cite{motwani2004survey, fan2019brief,ilesanmi2021methods} for a brief review. 

The problem of image denoising can be mathematically formulated as
\[
	\boldsymbol{y} = \boldsymbol{x}+ \boldsymbol{\epsilon},
\]
where $\boldsymbol{y}$ is the noised image,  $\boldsymbol{x}$ is the ground truth image and $\boldsymbol{\epsilon}$ is the noise. 
{Given the noised image $\boldsymbol{y}$, the task 
is to recover the ground truth $\boldsymbol{x}$ by eliminating the noise $\boldsymbol{\epsilon}$.}

In general, image denoising methods can be categorized into two main groups: classical methods and machine learning-based methods. Classical methods are rooted in mathematical models and signal processing techniques, such as 
median filtering \cite{farhang2013adaptive}, wavelet denoising \cite{chang2000adaptive}, total variation denoising \cite{rudin1992nonlinear},  non-local means filtering \cite{buades2005review}, to list a few. On the other hand, machine learning-based methods use artificial neural networks or other machine learning algorithms to learn the mapping between noisy and clean images. There are many machine learning-based models for image denoising developed recently which achieve dramatic success. 
Popular deep learning methods in this area include image denoising CNNs (DnCNN) \cite{zhang2017beyond}, residual dense networks (RDN) \cite{zhang2018residual}, non-local neural networks (N3Net) \cite{plotz2018neural}, Batch renormalization denoising network (BRDNet) \cite{tian2020image}, to list a few.


Among the classical methods, partial differential equations (PDE) provide a choice that utilizes evolving models to remove noise from digital images. The traditional PDE-based image denoising method did not utilize neural networks, including the Perona-Malik method \cite{perona1990scale}, Total Variation (TV) method \cite{rudin1992nonlinear}, Fast Marching Methods (FMM) \cite{sethian1996fast}, to list a few. 


While traditional PDE-based methods have many advantages including preserving edges and textures accurately, they also come with certain drawbacks. Specifically, using classical numerical methods to solve non-linear PDEs can lead to high computational intensity. Additionally, each fixed PDE model has limited capabilities in handling complex noise. These limitations can make traditional PDE-based methods less suitable for challenging applications, and researchers continue to explore alternative approaches to image denoising that may offer improved performance and versatility.


Recently several neural network methods have been proposed to learn PDE for various imaging tasks \cite{de2013image, ashouri2022new}.  One advantage of using learning-based PDEs over traditional PDEs is that they do not require human intervention, as they are learned directly from training data. In contrast, the traditional PDE approach often involves human input and expertise in formulating mathematical equations. In this article, we also take a similar learning-based approach. Such an approach allows us not to propose any specific dynamic system in the first instance, and instead learn it through a general ODE system modeled by the neural ODE (NODE) method \cite{chen2018neural}.

{
More precisely, we assume that the image denoising task can be modeled by the following dynamic system:
\begin{equation}\label{PDE-model}
 \boldsymbol{v}_t = F(t,\boldsymbol{v}), \quad \boldsymbol{v}(t=0) = \boldsymbol{y},
  \quad \boldsymbol{v}(t=T) = \boldsymbol{x},
\end{equation}
where the functional $F(t,\boldsymbol{v})$ is to be learned. Thanks to its generality, we note that $F$ can potentially represent an implicit differential operator involving spacial derivatives. Therefore, under such case \cref{PDE-model} represents an implicit PDE system. It is important to mention that we omit the direct dependence of the spacial variable, denoted by  $\boldsymbol{\tau}$, in $F(t,\boldsymbol{v})$, since we desire the spacial invariance property. In other words, one should obtain the same denoised result for a noised pixel regardless of its position in a man-made coordinate system.
}

{
Though \cref{PDE-model} represents a very general framework, it may be more effective to explicitly add $\nabla_{\boldsymbol{\tau}}\boldsymbol{v}$ into the input list of functional $F$ since $\nabla_{\boldsymbol{\tau}} \boldsymbol{v}$ provides ample useful information including edging, texture, etc. Here $\boldsymbol{\tau}$ represents the spacial variables.
 In this case, our model renders $\boldsymbol{v}_t = F(t,\boldsymbol{v}, \nabla_{\boldsymbol{\tau}} \boldsymbol{v})$ and becomes an explicit PDE system.  We will consider this as future work. 
}

{
Since our model  in \cref{PDE-model} represents a general ODE system, it is then natural to utilize a NODE-based neural network \cite{chen2018neural} to model the system. 
The NODE network has been exploited to efficiently approximate high-dimensional parabolic PDE problems in \cite{oliva2022towards}. 
NODE is also closely related to the Deep Residual Network (ResNet) \cite{he2016deep} as ResNets can be viewed as a discrete-time approximation of a Neural ODE. 
The authors in \cite{chen2018neural} demonstrate that NODEs are capable of diverse tasks, such as image classification, density estimation, and generative modeling, and show that they can outperform traditional neural networks on certain tasks while using fewer parameters.}

\subsection{Our contribution}
Inspired by the PDE method, we identify the image-denoising task as learning a general dynamical ODE system with an unknown vector field in \cref{{PDE-model}}. 
The main features of our proposed model include:
\begin{itemize}
\item It successfully captures the dynamics of the image denoising process and preserves intrinsic features of the images, such as edges and textures; 
    \item It allows flexible model choices to approximate the unknown vector field, thus is able to utilize/build on the state-of-the-art model(s) for imaging denoising;
    \item It is able to learn effectively from small amount of training imags without inducing overfitting issues;
    \item It admits flexible time steps in the ODE solver, which enable easy adjustment of the model complexity based on task needs. 
    \item It is capable of learning noise levels of a wide range, and, compared to baseline models, the benefit becomes more pronounced as the level of noise increases. 
\end{itemize}
Each of the points will be presented in Section \ref{sec:secproposed} with supporting evidence in Section \ref{sec:expertimetal}.



Our paper is organized as follows: Section \ref{sec:secproposed} is devoted to the detailed introduction of our proposed NODE-ImgNet model, where in Section \ref{sec:node} we also recall the structure of NODE model. In \Cref{sec:expertimetal}, we present the results of our proposed method on multiple benchmark datasets, which include two grayscale image datasets (BSD68 and Set12) and three color-scale image datasets (CBSD68 Kodak24 and McMaster). Our method achieves state-of-the-art performance in terms of peak signal-to-noise ratio (PSNR) across multiple scales of additive Gaussian noise. Moreover, our proposed model surpasses other highly competitive baseline models in real image denoising tasks. Discussion, limitation and future works are included in Section \ref{sec:discussion}.

\section{NODE-ImgNet for Image Denoising }\label{sec:secproposed}
\subsection{NODE}\label{sec:node}
We first briefly recall the NODE model.
NODE models a continuous function that maps $\mathbb{R}^{d}$ to $\mathbb{R}^{d}$ 
by defining the ODE model as follows: for every input $\mathbf{h}_0 \in \mathbb{R}^{d}$, the corresponding output of NODE model is $\textbf{h}(T)$ (or $\textbf{h}(t)_{t \in [0, T]}$) where $\textbf{h}$ satisfies 
 \begin{eqnarray}\label{NODE}
\frac{\mathrm{d}\textbf{h}(t)}{\mathrm{d}t}=\mathcal{N}^{\text{vec}}_{\boldsymbol{\theta}}(\textbf{h}(t), t), \quad \mathbf{h}(0) = \mathbf{h}_0.
\end{eqnarray}
Here the vector field  $\mathcal{N}^{\text{vec}}_{\boldsymbol{\theta}}$ represents a neural network with trainable parameters $\boldsymbol{\theta}$.
In the image denoising model, it is natural to take $ \mathbf{h}_0$ as the initial noisy image $\mathbf{y}$ and $ \mathbf{h}(T)$ the denoised image, ideally $\mathbf{x}$.
By Picard's Theorem, the initial value problem \eqref{NODE} yields a unique solution provided $\mathcal{N}_{\boldsymbol{\theta}}^{\text{vec}}$ is Lipschitz continuous, which is guaranteed by using finite weights and activation functions like $\texttt{relu}$ or $\texttt{tanh}$ \cite{chen2018neural}.

The NODE model can be viewed as a continuous and more accurate version of the deep residual learning neural network (Res-Net) \cite{he2016deep}. The element of residual blocks have been frequently used in many state-of-art image-denoising neural networks, e.g., DnCNN \cite{zhang2017beyond},FFDNet \cite{zhang2018ffdnet}, DudeNet \cite{tian2021designing} and batch-renormalization denoising network (BRDNet) \cite{tian2020image}. 

It is important to note that we have the flexibility to utilize various state-of-art image denoising network models for the vector field $\mathcal{N}^{\text{vec}}_{\boldsymbol{\theta}}$.


With the recent increase in computational power, state-of-the-art image denoising models have become more complex, with millions of tunable parameters \cite{park2019densely,yu2019deep,gurrola2021residual}.  
However, this increased complexity often requires a large amount of training data to achieve optimal results. In many applications where images are obtained at a high cost, there may be insufficient training data, leading to over-fitting with large iterations. One of our aim is to enable our network capable to handle different training data sizes, providing flexibility. Our guidance is to use a simpler model for smaller training data sets and a more complex model for larger ones through a simple adjustment of time integration steps used in the ODE numerical solver.

More specifically, the complexity of our NODE-based neural network arises from two factors: the vector field $\mathcal{N}^{\text{vec}}_{\boldsymbol{\theta}}(\textbf{h}(t), t)$ on the right-hand side and the time integration number $N$ used in the numerical ODE solver. As aforementioned, we have the flexibility to construct our vector network $\mathcal{N}^{\text{vec}}_{\boldsymbol{\theta}}(\textbf{h}(t), t)$ using many state-of-the-art networks as a basis. Notably, setting the step size $N$ to $1$ 
results in the NODE model degenerating into the vector network. Increasing the time steps introduces additional time integration steps, similar to residual blocks.


However, unlike classical residual blocks, increasing the time step in the NODE network does not increase the number of model parameters; only the training time increases linearly. To balance the two components of complexity, we propose using a structure-similar yet much simpler analogue of state-of-the-art neural networks for the vector field. The model will then tune an appropriate step size to compensate for its accuracy. Since our model uses fewer parameters, it tends to be more effective for small training sets. Furthermore, by increasing the time step, we can increase the model's complexity, making it also effective for larger training sets.

\subsection{Our proposed NODE-ImgNet} \label{sec:proposed}
 We use  \eqref{NODE} to model the dynamics of the image denoising process. In particular, we choose to use a shallow CNN with batch normalization and dilation as our \emph{vector field} for image denoising. Such structure has been used as a basic block in many state of art networks for image denoising task, e.g., the BRDNet \cite{tian2020image}, DilatedConv-BN-ReLU \cite{wang2017dilated}, FFDNet \cite{zhang2018ffdnet}, DudeNet \cite{tian2021designing}, etc.
 

\begin{figure}[h!]
\centering
\includegraphics[width=1\textwidth]{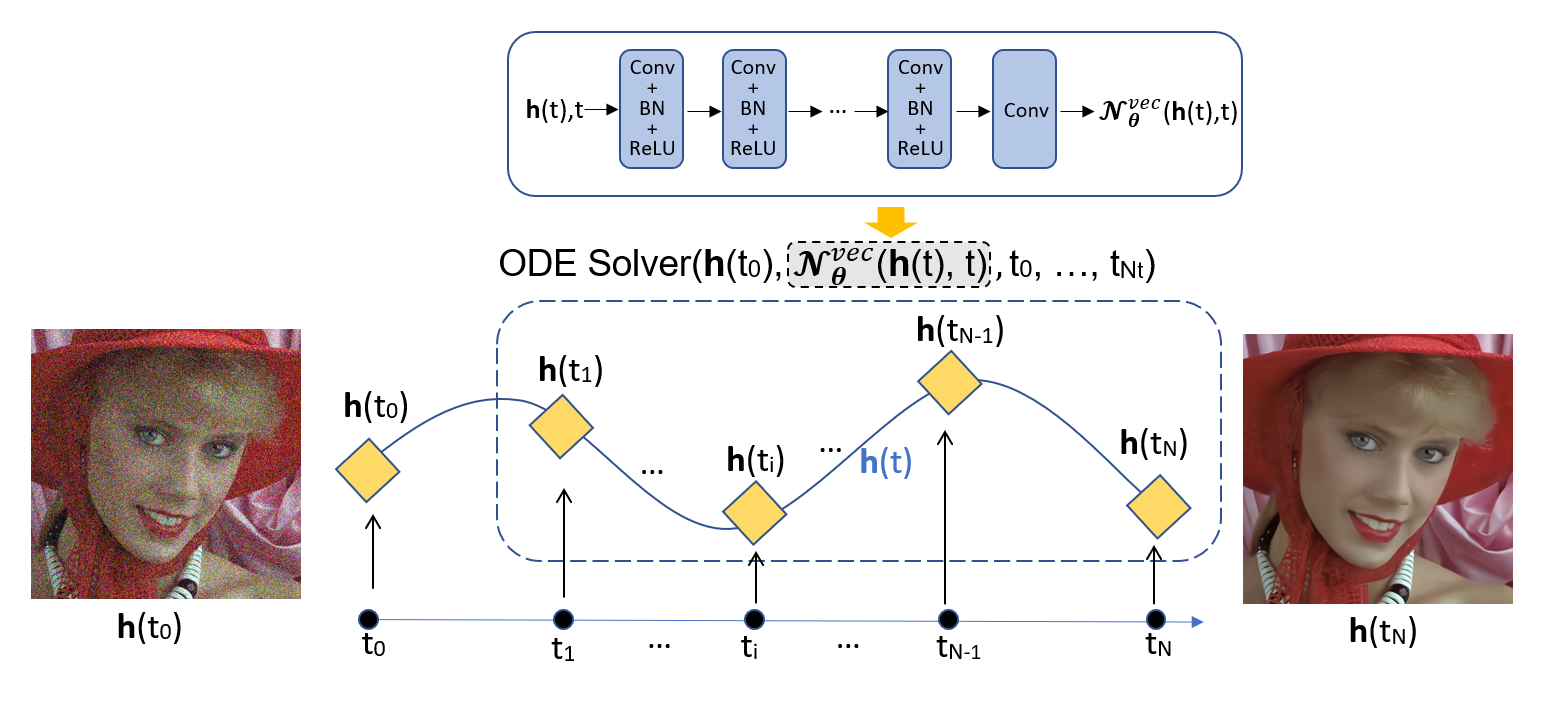}
\caption{Architecture of proposed NODE-ImgNet network.}
\label{pic:structure}
\end{figure}

The structure of the shallow CNN vector field consists of a 9-layer CNN, with integrated batch normalization (BN) and dilated convolution. Batch normalization and dilated convolution have been proven effective and widely used in the construction of Deep CNN models \cite{wang2017dilated,ioffe2015batch}. BN has been proven effective in removing exploding or vanishing gradients, as well as resolving the internal covariate shift problem caused by convolutional operations \cite{ioffe2015batch}. The dilation factor and the number of dilated convolution layers directly determine the receptive field size of the network. We refer \Cref{pic:structure} for a graphic view of NODE-ImgNet.

The detailed structure of the $9\text{-layer}$ vector field is comprised of two types of structures: $\text{Conv}+\text{BN}+\text{ReLU}$ and Conv. The term `Conv' refers to a dilated convolutional layer with $3\times3$ filters and ReLU refers to rectified linear units.

Layers $1$-$8$ are of the $\text{Conv}+\text{BN}+\text{ReLU}$ type and the $9$-th layer is Conv. For all layers except the first and the last, the in-channel and out-channel numbers are both set to $128$, unless otherwise specified. 
The in-channel number of the first layer is set to $c + 1$ where $c$ is the channel of the image with the extra channel being the time variable.  For the final layer,  the out-channel number is set to be $c$.

For the first and last two layers, the dilation factor is set to be $1$, which is equivalent to the normal convolution without dilation.
The dilation factor of $4$ is applied in the convolution layers $2$-$7$. 
Applying such dilated convolutions gradually increases the receptive field size of the network. 
More precisely, the initial receptive field size is $3$ in the first layer, the same as the kernel size. For each subsequent layer, the size of the receptive field is increased by {$(\text{kernel size} - 1) \times \text{dilation factor}$}. Therefore, the receptive field sizes of our model in the nine layers are $3, 11, 19, 27, 35, 43, 51, 53$ and $55$, respectively. 
We note that the dilation factors are chosen such that
final receptive field size is close to the size of the image patches. 


Recall that one of the key features of our model is the flexibility to control its depth by adjusting the number of time integration steps, $N$. In the experiments described in the paper, we set $N$ to be between $2$ and $8$, which allowed us to balance the training time and the expected performance of the model effectively. 
For the numerical integration, we use the randomized (forward) Euler method \cite{kruse2017error}. Finally, we choose the final time parameter $T=1$.

\subsection{Loss function}
In our work, 
we choose the mean-square error (MSE). 
Define
\begin{equation}
   l(\boldsymbol{\theta}) = \frac{1}{N_p}\sum ^{N_p}_{j=1}||\boldsymbol{f}(\boldsymbol{y_i},\boldsymbol{\theta})-\boldsymbol{x_i}||_2^2, 
   \label{equ:lossfuction}
\end{equation}
where $N_p$ is the number of noisy-image patches, $\boldsymbol{\theta}$ denotes the set of all parameters in the model,  $f(\boldsymbol{y_i},\boldsymbol{\theta})$ denotes the output of the neural network at the final time $t=T$ for the denoised image $\boldsymbol{y_i}$, i.e.,
 $\boldsymbol{f}(\boldsymbol{y_i},\boldsymbol{\theta}) = \boldsymbol{h}(T)$
 with $\boldsymbol{h}(0) = \boldsymbol{y_i}$ in \eqref{NODE}.
$\boldsymbol{x}_i$ is the ground truth image corresponding to $\boldsymbol{y_i}$.
The norm $\|\cdot\|_2$ denotes the Euclidean norm.
\section{Experimental results}\label{sec:expertimetal}
{
This section mainly presents experimental results of our NODE-ImgNet model along with comparisons to other competitive baseline models that have similar CNN structures. The datasets and experimental settings used in the experiments are documented in Section \ref{sec:datasets} and \ref{sec:Expaermental setting}. We compare the model performance on the synthetic Gaussian noise denoising for both gray and color images in Section \ref{sec:gray Gaussian} and \ref{sec:color Gaussian} respectively. We also perform experiments on the real images and the results are presented in Section \ref{sec:real noise}. {After that, we investigate the scalability of our proposed method on small-scale partial datasets in Section \ref{advantage1}. In Section \ref{sec:flexibility}, we present our ablation experiments and demonstrate the flexibility of our model. Lastly, we discuss the computational costs in the subsequent Section \ref{sec:computational costs}.}
The codes are avaliable at \url{https://github.com/xiexinheng/NODE-ImgNet}. 
} 
 
\subsection{Datasets}\label{sec:datasets}
\subsubsection{Training dataset}
For the experiments of Gaussian noise denoising, we train our model on the Waterloo Exploration Database, which contains $4744$ color images. 
The full database is used for both color and gray image denoising tasks, where for the latter we first convert the original color images to gray images. 
 To test the effectivity of our model in handling data of various sizes, we also train the model based on $2\%$, $10\%$, $20\%$, and $50\%$ portion of the full image set with random sampling.
In all experiments, each image from the dataset is cropped into $300$ small $60\times60$ sized patches. The choice of $60\times60$ as the patch size is made as a trade-off between computational efficiency and performance. {The Gaussian noises are added as independent and identically distributed (i.i.d.) random variables with a mean of $0$ and a standard deviation of $\sigma$ (either fixed or to be specified in a range) to each patch in the training dataset.}


{For the real image denoising, we train our model based on a dataset of $40$ image pairs collected by Xu et al. \cite{xu2018real}, as well as the SIDD Medium dataset \cite{SIDD_2018_CVPR}, which includes $320$ image pairs. Each image pair contains a ground truth image and a corresponding noisy image.} Since the images in the first dataset are relatively small, we augment the dataset by applying classical image augmentation techniques such as rotations of $90$°, $180$°, and $270$°, resulting in $120$ additional images. We also apply horizontal flipping to each of the $160$ images, resulting in an additional $160$ images. Combining the augment data set with the SIDD dataset, we obtain a total of $640$ images. Finally, each image is cropped into $662$ patches of size $60\times 60$ pixels, resulting in a total of $423,680$ training patches.


\subsubsection{Test dataset}
For a fair comparison, we follow other Gaussian image denoising models 
\cite{zhang2017beyond, zhang2018ffdnet, tian2020image} and test our model on datasets BSD68 and Set12 for the trained gray-scale image models, and on CBSD68, Kodak24 and McMaster for the trained color image models. 
BSD68 consists of $68$ natural images with dimensions of either $481 \times 321$ or $321 \times 481$; Set12 consists of $12$ gray-scale images with a size of either $256 \times 256$ or $512 \times 512$. CBSD68 includes $68$ color images with the same background as BSD68. Kodak24 is composed of $24$ natural color images, each with a size of $500 \times 500$. McMaster consists of $18$ color images, each with a size of $500 \times 500$.

\begin{figure}[h!]
\centering
\includegraphics[width=1\textwidth]{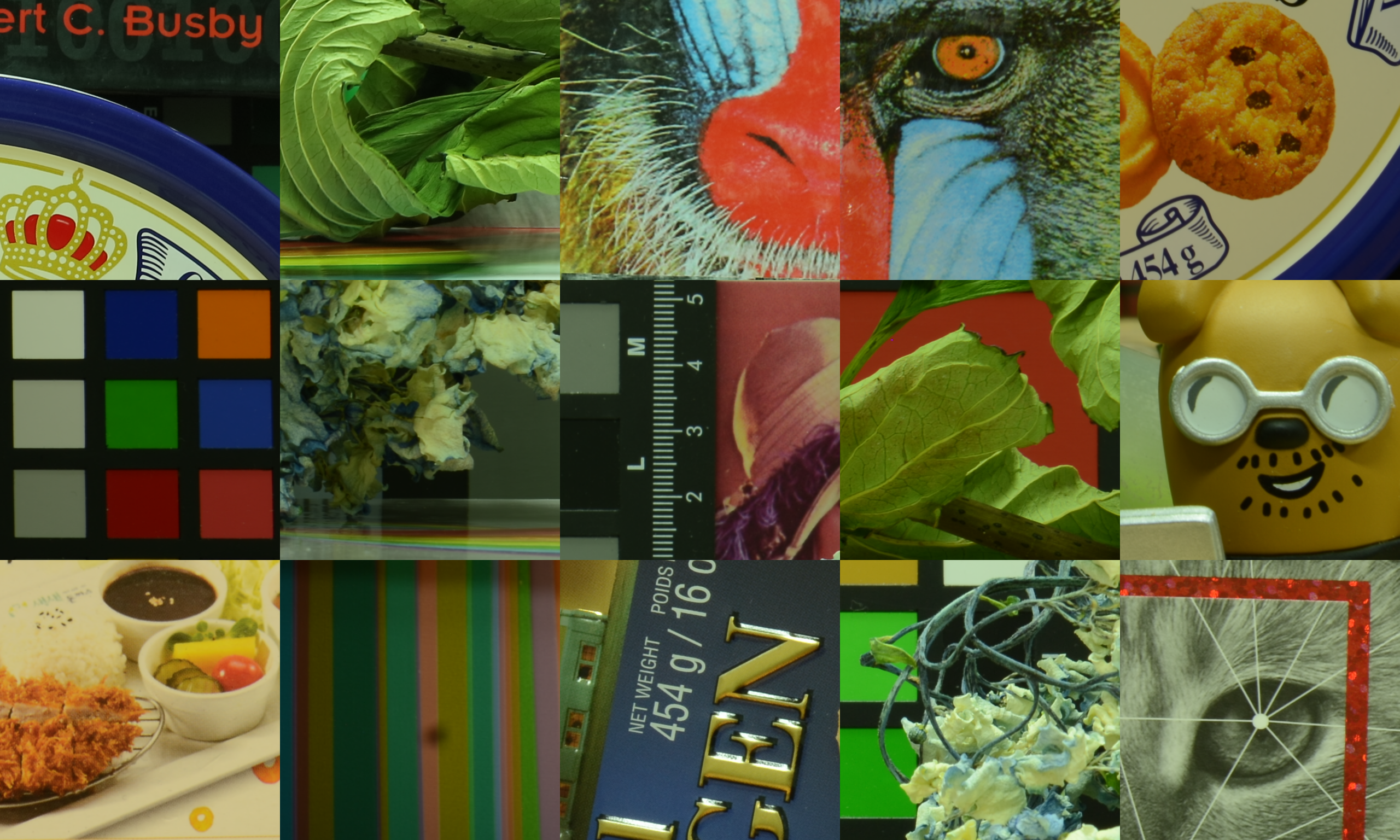}
\caption{$15$ images in the cc dataset.}
\label{test_images}
\end{figure}

It is known that real noisy images are usually captured by cameras of different types with different ISO values \cite{dabov2007color}. Motivated by this fact, we choose cc \cite{nam2016holistic} as the test dataset for real noise task. The cc dataset has $15$ noisy images captured by three different cameras, i.e., Nikon D800, Nikon D600, and Canon 5D Mark III with different ISO values ($1600$, $3200$, and $6400$), shown in \Cref{test_images}. 

\subsection{Other  experimental settings}
\label{sec:Expaermental setting}
{We use the Adam optimizer with a learning rate of $5\times 10^{-4}$, $\beta_1$ of $0.9$, $\beta_2$ of $0.999$, and epsilon of $1\times 10^{-8}$. Here, $\beta_1$ and $\beta_2$ control the exponential decay rates for the first and second moment estimates of the gradient, respectively, while $\epsilon$ is a small value added to the denominator of the Adam update formula to prevent division by zero.} The batch size is set to $40$, and the maximum number of epochs for training NODE-ImgNet is $50$. {During the training, the learning rate is halved if the training loss does not decrease for three consecutive epochs. 
Additionally, we employ early stopping techniques to reduce training time and avoid overfitting. 
In particular, the loss on a specified test set is computed after each epoch. For gray and color image denoising tasks, we use the set68 and McMaster  test sets, respectively. The early stopping is activated if the loss of test set does not decrease for five consecutive epochs.}

We train our proposed NODE-ImgNet model under the PyTorch framework \cite{NEURIPS2019_9015}. All experiments are conducted on a server running the Linux-$5.4.0$ operating system and Python $3.9$ environment. The server has a memory capacity of $1.4$ TiB and is powered by an Intel Xeon(R) Gold 6240 CPU @ $2.60$GHz x $72$G processor. The server is equipped with an NVIDIA GeForce RTX $3080$ Ti GPU, which has $12$ GB of GPU RAM. To accelerate the computational performance of the GPU, we use the NVIDIA CUDA $11.8$ version. We use the $0.2.3$ version of the torchdiffeq package to solve NODEs. This package is optimized for GPU usage and offers a way to reduce memory consumption through backpropagation. 
{To provide the reader with a point of reference, the training of a color Gaussian denoising NODE-ImgNet model, employing the structure presented in Section \ref{sec:proposed} and utilizing $N=8$, requires an estimated duration of around $82$ hours.} 

\subsection{Compared results}\label{sec:compariosn}
In this section, we will give both quantitative and qualitative analyses of the performance by NODE-ImgNet. 
We compare NODE-ImgNet with other competitive baseline image denoising methods, including DnCNN \cite{zhang2017beyond}, Dilated Conv-BN-ReLU(DC-B-R) \cite{wang2017dilated}, FFDNet \cite{zhang2018ffdnet}, complex-valued denoising network (CDNet) \cite{quan2021image}, DudeNet \cite{tian2021designing}, batch renormalization denoising network(BRDNet) \cite{tian2020image}, ADNet \cite{tian2020attention}, MWDCNN \cite{tian2023multi} and GradNet \cite{liu2020gradnet}.
We select these models because they are primarily constructed using Convolution (Conv), Batch Normalization (BN), Residual Learning (RL), and Rectified Linear Unit (ReLU) technologies, or their variations such as complex-valued and batch renormalization. Comparing our model to these existing models allows for a clear demonstration of the improvements made by our model. 

In this section, we will employ $N=8$ by default as the number of time integration steps in the ODE solver for NODE-ImgNet models, unless specified otherwise. This specific hyperparameter selection is based on finding the optimal balance between accuracy and training cost, ensuring an effective trade-off between the two.

In the following, we first present a comparison of NODE-ImgNet's denoising performance on gray images using the BSD68 and Set12 test sets in \Cref{sec:gray Gaussian}. Next, in Section \ref{sec:color Gaussian}, we evaluate the denoising capabilities of NODE-ImgNet on color images of three test datasets: CBSD68, Kodak24, and McMaster. We also compare NODE-ImgNet's denoising ability on real images using the cc dataset in Section \ref{sec:real noise}. 

\subsubsection{NODE-ImgNet for gray Gaussian image 
denoising}\label{sec:gray Gaussian}
For gray Gaussian image denoising, we train
multiple NODE-ImgNet models in two modes. In the first mode, following literature \cite{zhang2017beyond, zhang2018ffdnet, tian2020image} we train three separate models  with fixed noise level $\sigma = 15, 25, 50$, respectively. We refer to these three models as NODE-ImgNet. 
In the second mode, we employ a blind denoising approach, where 
the noise level $\sigma$ for each image patch is randomly sampled based on a uniform distribution between $0$ and $55$ during the training process.
We refer to this model as NODE-ImgNet-B, which aimed to provide a robust image denoising model that can deal with a wide range of noise levels in real-world scenarios.

{
The test results of our trained models along with comparisons based on the BSD68 dataset are presented in Table \ref{tab:gray_BSD68}, where the highest and second-highest PSNR values for each noise level are highlighted in red and blue, respectively. The results show that NODE-ImgNet outperforms all the compared benchmark models in terms of PSNR. 
Moreover, we observe a more significant improvement in our model as the noise level increases.
{In particular, compared to the baseline model DnCNN, whose structure is similar to the vector field of NODE-ImgNet, NODE-ImgNet achieves average improvements of $0.09, 0.13, 0.22$ (in dB)  for $\sigma=15, \,25$ and $50$, respectively.}
NODE-ImgNet-B also shows competitive performance. Notably, when test $\sigma=50$, NODE-ImgNet-B performs better than all other models except NODE-ImgNet. 


\renewcommand{\arraystretch}{1.5}

\begin{table}[]
\resizebox{\columnwidth}{!}{%
\begin{tabular}{|c|c|c|c|c|c|c|c|c|c|c|}
\hline
\backslashbox{\hspace{12mm}$\sigma$}{Methods}      & \textbf{\makecell[c]{DnCNN \\ \cite{zhang2017beyond}} } & \textbf{\makecell[c]{DC-B-R\\ \cite{wang2017dilated}}} & \textbf{\makecell[c]{FFDNet\\ \cite{zhang2018ffdnet}}} & \textbf{\makecell[c]{ADNet\\\cite{tian2020attention}}} & \textbf{\makecell[c]{DudeNet\\ \cite{tian2021designing}}} & \textbf{\makecell[c]{CDNet\\\cite{quan2021image}}} & {\color[HTML]{2E2E2E} \textbf{\makecell[c]{MWDCNN\\\cite{tian2023multi}  }}} & \textbf{\makecell[c]{BRDNet\\\cite{tian2020image}}   } & \textbf{\makecell[c]{NODE-\\ImgNet}}         & \textbf{\makecell[c]{NODE-\\ImgNet-B}}       \\ \hline
\textbf{$\sigma =15$} & 31.72                                & 31.68                                              & 31.63                                 & 31.74                                  & 31.78                                    & 31.74                              & 31.77                                                         & {\color[HTML]{0070C0} 31.79}           & {\color[HTML]{C00000} 31.81} & 31.66                        \\ \hline
\textbf{$\sigma =25$} & 29.23                                & 29.18                                              & 29.19                                 & 29.25                                  & 29.29                                    & 29.28                              & 29.28                                                         & {\color[HTML]{0070C0} 29.29}           & {\color[HTML]{C00000} 29.36} & 29.27                        \\ \hline
\textbf{$\sigma =50$} & 26.23                                & 26.21                                              & 26.29                                 & 26.29                                  & 26.31                                    & 26.36                              & 26.29                                                         & 26.36                                  & {\color[HTML]{C00000} 26.45 } & {\color[HTML]{0070C0} 26.41} \\ \hline
\end{tabular}%
}
\caption{Average PSNR (dB) results of various methods for Gaussian gray image denoising on the BSD68 Dataset at various noise levels.}
\label{tab:gray_BSD68}
\end{table}
   
In \Cref{pic:set68-sigma25}, we compare the results through visualization of a same zoomed region  
from denoised images produced by various methods when $\sigma=25$. 
We observe that the denoised image generated by NODE-ImgNet exhibits better quality in terms of preserving edges and detailed textures.


\begin{figure}[h]
    \centering
    \begin{subfigure}{0.32\textwidth}
        \includegraphics[width=\textwidth]{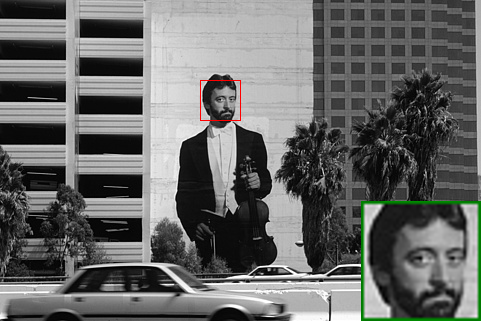}
        \caption{Original Image}
        \label{fig:set68 image1}
    \end{subfigure}
    \hspace{0em} 
    \begin{subfigure}{0.32\textwidth}
        \includegraphics[width=\textwidth]{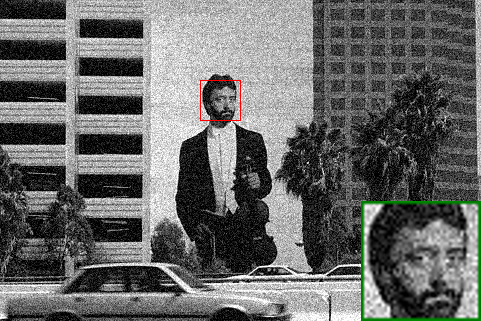}
        \caption{Noisy Image ($20.71$dB)}
        \label{fig:set68 image2}
    \end{subfigure}
    \hspace{0em} 
    \begin{subfigure}{0.32\textwidth}
        \includegraphics[width=\textwidth]{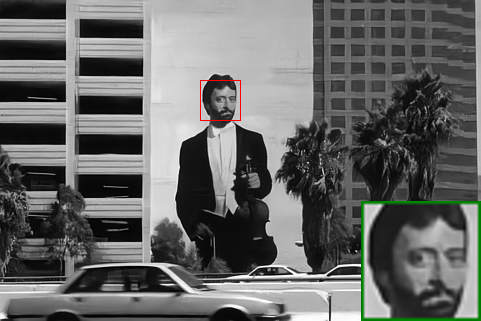}
        \caption{DnCNN ($29.17$dB)}
        \label{fig:set68 image3}
    \end{subfigure}

    \vspace{0.3em}
    \begin{subfigure}{0.32\textwidth}
        \includegraphics[width=\textwidth]{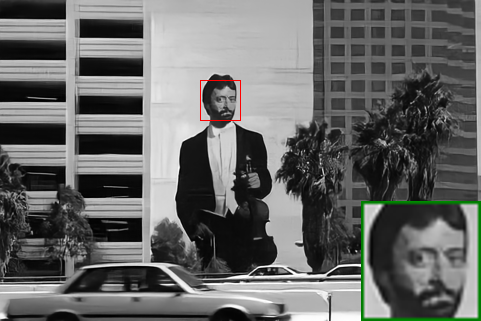}
        \caption{ffdnet ($29.15$dB)}
        \label{fig:set68 image4}
    \end{subfigure}
    \hspace{0em} 
    \begin{subfigure}{0.32\textwidth}
        \includegraphics[width=\textwidth]{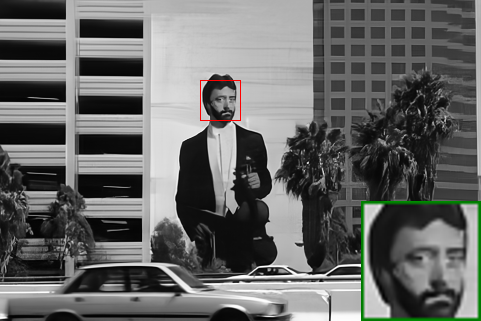}
        \caption{BRDNet ($29.51$dB)}
        \label{fig:set68 image5}
    \end{subfigure}
    \hspace{0em} 
    \begin{subfigure}{0.32\textwidth}
        
        \includegraphics[width=\textwidth]{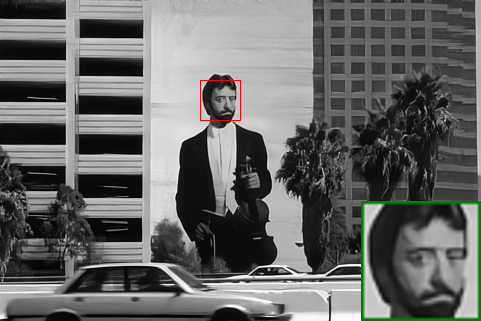}
        \caption{ MWDCNN ($29.25$dB)}
        \label{fig:set68 image6-1}
    \end{subfigure}
    \vspace{0.3em}

    \begin{subfigure}{0.32\textwidth}
    \includegraphics[width=\textwidth]{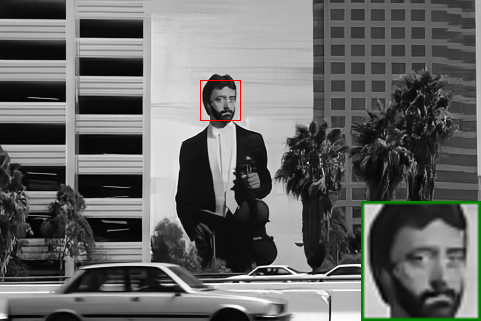}
        \caption{NODE-ImgNet-B ($29.52$dB)}
        \label{fig:set68 image6}
    \end{subfigure}
        \hspace{0em} 
    \begin{subfigure}{0.32\textwidth}
        \includegraphics[width=\textwidth]{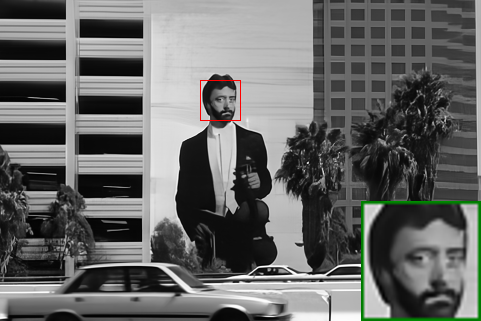}
        \caption{NODE-ImgNet ($29.56$dB)}
        \label{fig:set68 image7}
    \end{subfigure}
\caption{Denoising results for Gaussian gray noisy image from the Set68 dataset with $\sigma = 25$. }
\label{pic:set68-sigma25}
\end{figure}

\Cref{pic:set12} visualise all test images from Set12. In \Cref{tab:gray_set12}, 
we present the  comparison results for each individual image from Set12 on various noise levels.
As before, the highest and second-highest PSNR results for each setting are marked in red and blue, respectively.
We again observe that NODE-ImgNet on average outperforms all other models. In particular, compared to the baseline model DnCNN, NODE-ImgNet achieves average improvements of $0.18, 0.30, 0.45$ (in dB)  for $\sigma=15, 25$ and $50$, respectively.
Clearly, NODE-ImgNet demonstrates superior performance compared to other models across a wide range of image settings. In particular, when evaluated on the Set12 dataset with noise levels of $\sigma=15, 25$, and $50$, NODE-ImgNet achieves the highest PSNR results for $58\%$, $71\%$, and $92\%$ of the images, respectively.
} 
Similar to previous observations, as the noise level increases, our model achieves more significant improvement.

\begin{figure}[hbt!]
    \centering
    \begin{subfigure}{0.32\textwidth}
        \includegraphics[width=\textwidth]{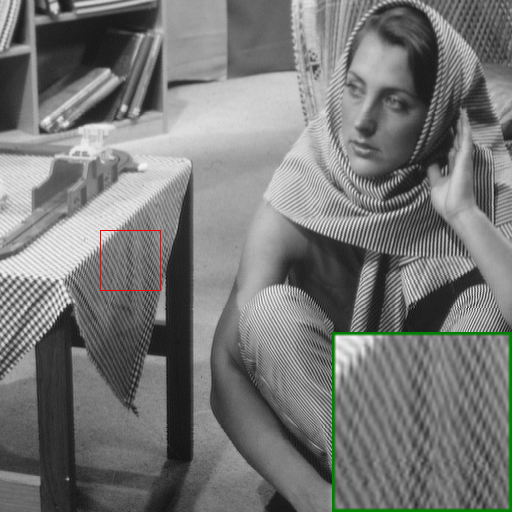}
        \caption{Original image}
        \label{fig:set12 image1}
    \end{subfigure}
    \hspace{0em} 
    \begin{subfigure}{0.32\textwidth}
    \includegraphics[width=\textwidth]{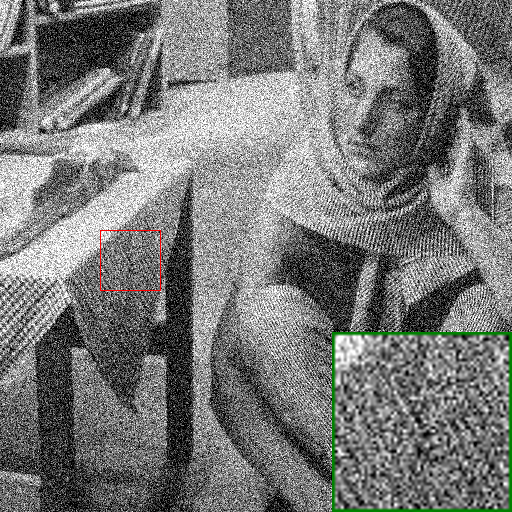}
        \caption{Noisy image ($14.77$dB)}
        \label{fig:set12 image2}
    \end{subfigure}\\
    \begin{subfigure}{0.32\textwidth}
        \includegraphics[width=\textwidth]{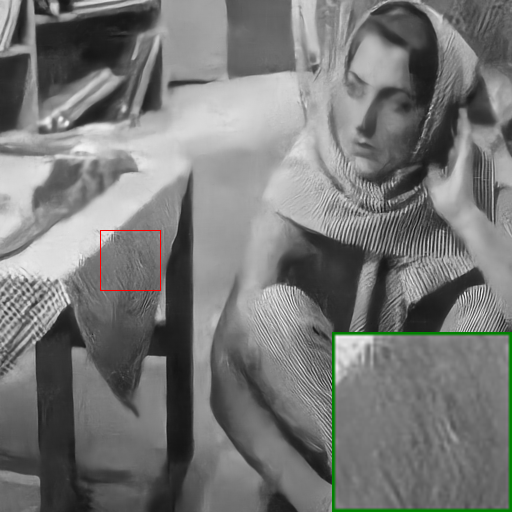}
        \caption{DnCNN ($26.22$dB)}
        \label{fig:set12 image3}
    \end{subfigure}
   \hspace{0.3em}
    \begin{subfigure}{0.32\textwidth}
        \includegraphics[width=\textwidth]{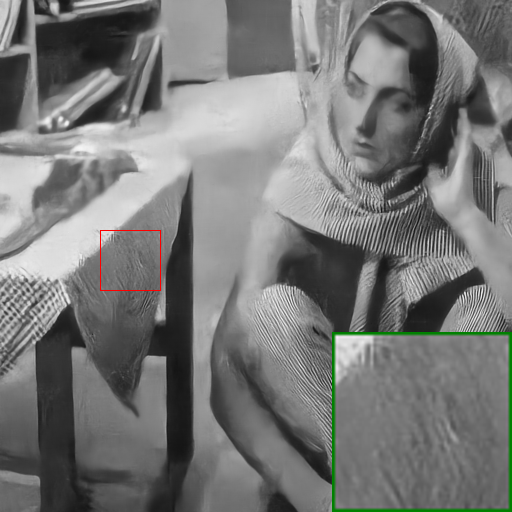}
        \caption{FFDNet ($26.45$dB)}
        \label{fig:set12 image4}
    \end{subfigure}
    \hspace{0em} 
    \begin{subfigure}{0.32\textwidth}
        \includegraphics[width=\textwidth]{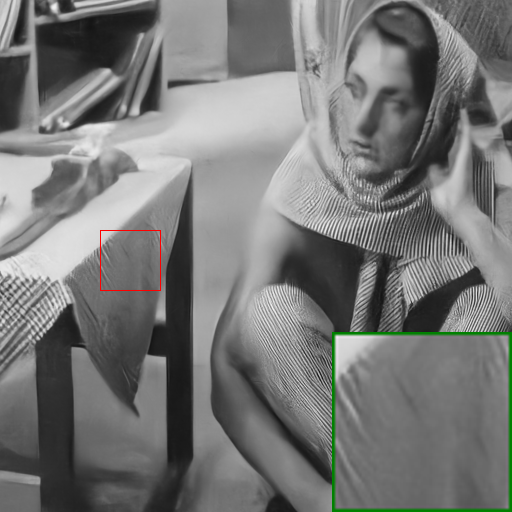}
        \caption{BRDNet ($26.85$dB)}
        \label{fig:set12 image5}
    \end{subfigure}
    \hspace{0em} 
    \begin{subfigure}{0.32\textwidth}
    \includegraphics[width=\textwidth]{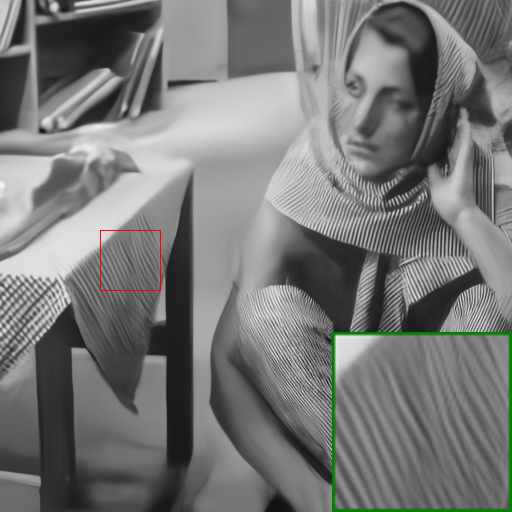}
        \caption{NODE-ImgNet-B ($27.08$dB)}
        \label{fig:set12 image6}
    \end{subfigure}
   \hspace{0em} 
    \begin{subfigure}{0.32\textwidth}
        \includegraphics[width=\textwidth]{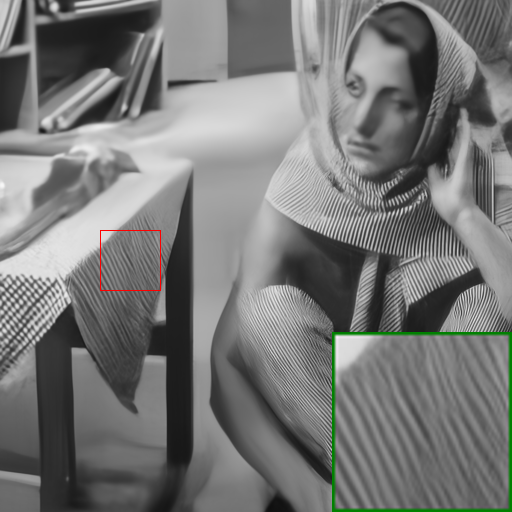}
        \caption{NODE-ImgNet ($27.32$dB)}
        \label{fig:set12 image7}
    \end{subfigure}
\caption{Denoising results for Gaussian gray noisy image “Barbara” from the Set12 dataset with $\sigma = 50$.}
\label{pic:set12-visual}
\end{figure}

In \Cref{pic:set12-visual}, we compare the results through visualization of the same zoomed region  
from denoised images produced by various methods when $\sigma=50$. It is obvious that NODE-ImgNet is better at preserving edges and complex textures.

\begin{figure}[hbt!]
\centering
\includegraphics[width=1\textwidth]{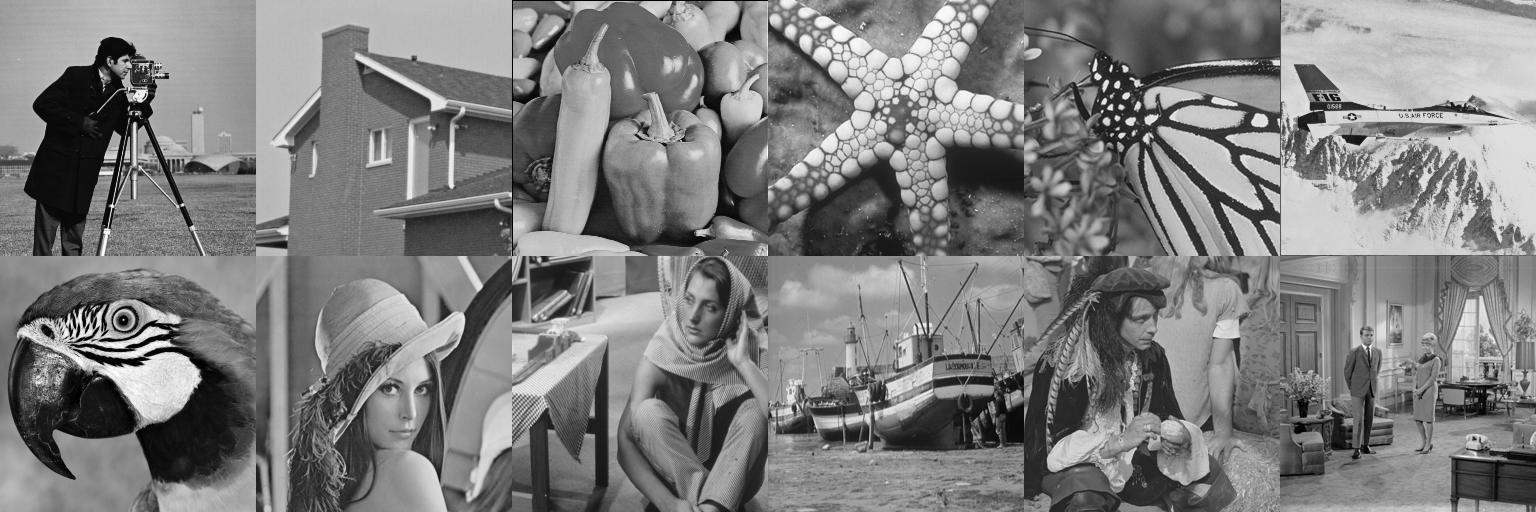}
\caption{The $12$ images in the Set12 dataset.}
\label{pic:set12}
\end{figure}

\begin{table}[hbt!]
\resizebox{\columnwidth}{!}{%
\begin{tabular}{|cccccccccccccc|}
\hline
\multicolumn{1}{|c|}{\backslashbox{Methods}{Images}}                          & \multicolumn{1}{c|}{\textbf{C.man}}               & \multicolumn{1}{c|}{\textbf{House}}               & \multicolumn{1}{c|}{\textbf{Peppers}}             & \multicolumn{1}{c|}{\textbf{Starfish}}            & \multicolumn{1}{c|}{\textbf{Monarch}}             & \multicolumn{1}{c|}{\textbf{Airplane}}            & \multicolumn{1}{c|}{\textbf{Parrot}}              & \multicolumn{1}{c|}{\textbf{Lena}}                & \multicolumn{1}{c|}{\textbf{Barbara}}             & \multicolumn{1}{c|}{\textbf{Boat}}                & \multicolumn{1}{c|}{\textbf{Man}}                 & \multicolumn{1}{c|}{\textbf{Couple}}              & \textbf{Average}             \\ \hline
\multicolumn{14}{|c|}{\textbf{Noise Level $\sigma= 15$}}                                                                                                                                                                                                                                                                                                                                                                                                                                                                                                                                                                                                                                                                                              \\ \hline
\multicolumn{1}{|c|}{\textbf{DnCNN \cite{zhang2017beyond}}}     & \multicolumn{1}{c|}{32.61}                        & \multicolumn{1}{c|}{34.97}                        & \multicolumn{1}{c|}{33.30}                        & \multicolumn{1}{c|}{32.20}                        & \multicolumn{1}{c|}{33.09}                        & \multicolumn{1}{c|}{31.70}                        & \multicolumn{1}{c|}{31.83}                        & \multicolumn{1}{c|}{34.62}                        & \multicolumn{1}{c|}{32.64}                        & \multicolumn{1}{c|}{32.42}                        & \multicolumn{1}{c|}{32.46}                        & \multicolumn{1}{c|}{32.47}                        & 32.86                        \\ \hline
\multicolumn{1}{|c|}{\textbf{FFDNet \cite{zhang2018ffdnet}}}    & \multicolumn{1}{c|}{32.43}                        & \multicolumn{1}{c|}{35.07}                        & \multicolumn{1}{c|}{33.25}                        & \multicolumn{1}{c|}{31.99}                        & \multicolumn{1}{c|}{32.66}                        & \multicolumn{1}{c|}{31.57}                        & \multicolumn{1}{c|}{31.81}                        & \multicolumn{1}{c|}{34.62}                        & \multicolumn{1}{c|}{32.54}                        & \multicolumn{1}{c|}{32.38}                        & \multicolumn{1}{c|}{32.41}                        & \multicolumn{1}{c|}{32.46}                        & 32.77                        \\ \hline
\multicolumn{1}{|c|}{\textbf{ADNet \cite{tian2020attention}}}   & \multicolumn{1}{c|}{32.81}                        & \multicolumn{1}{c|}{35.22}                        & \multicolumn{1}{c|}{33.49}                        & \multicolumn{1}{c|}{32.17}                        & \multicolumn{1}{c|}{33.17}                        & \multicolumn{1}{c|}{31.86}                        & \multicolumn{1}{c|}{31.96}                        & \multicolumn{1}{c|}{34.71}                        & \multicolumn{1}{c|}{32.80}                        & \multicolumn{1}{c|}{32.57}                        & \multicolumn{1}{c|}{32.47}                        & \multicolumn{1}{c|}{32.58}                        & 32.98                        \\ \hline
\multicolumn{1}{|c|}{\textbf{DudeNet \cite{tian2021designing}}} & \multicolumn{1}{c|}{32.71}                        & \multicolumn{1}{c|}{35.13}                        & \multicolumn{1}{c|}{33.38}                        & \multicolumn{1}{c|}{32.29}                        & \multicolumn{1}{c|}{33.28}                        & \multicolumn{1}{c|}{31.78}                        & \multicolumn{1}{c|}{31.93}                        & \multicolumn{1}{c|}{34.66}                        & \multicolumn{1}{c|}{32.73}                        & \multicolumn{1}{c|}{32.46}                        & \multicolumn{1}{c|}{32.46}                        & \multicolumn{1}{c|}{32.49}                        & 32.94                        \\ \hline
\multicolumn{1}{|c|}{\textbf{MWDCNN \cite{tian2023multi}}}     & \multicolumn{1}{c|}{32.53}                        & \multicolumn{1}{c|}{35.09}                        & \multicolumn{1}{c|}{33.29}                        & \multicolumn{1}{c|}{{\color[HTML]{0070C0} 32.28}} & \multicolumn{1}{c|}{33.20}                        & \multicolumn{1}{c|}{31.74}                        & \multicolumn{1}{c|}{31.97}                        & \multicolumn{1}{c|}{34.64}                        & \multicolumn{1}{c|}{32.65}                        & \multicolumn{1}{c|}{32.49}                        & \multicolumn{1}{c|}{32.46}                        & \multicolumn{1}{c|}{32.52}                        & 32.91                        \\ \hline
\multicolumn{1}{|c|}{\textbf{BRDNet   \cite{tian2020image}}}   & \multicolumn{1}{c|}{{\color[HTML]{C00000} 32.80}} & \multicolumn{1}{c|}{{\color[HTML]{0070C0} 35.27}} & \multicolumn{1}{c|}{{\color[HTML]{0070C0} 33.47}} & \multicolumn{1}{c|}{32.24}                        & \multicolumn{1}{c|}{{\color[HTML]{C00000} 33.35}} & \multicolumn{1}{c|}{{\color[HTML]{0070C0} 31.85}} & \multicolumn{1}{c|}{{\color[HTML]{C00000} 32.00}} & \multicolumn{1}{c|}{{\color[HTML]{0070C0} 34.75}} & \multicolumn{1}{c|}{{\color[HTML]{C00000} 32.93}} & \multicolumn{1}{c|}{{\color[HTML]{C00000} 32.55}} & \multicolumn{1}{c|}{{\color[HTML]{0070C0} 32.50}} & \multicolumn{1}{c|}{{\color[HTML]{0070C0} 32.62}} & {\color[HTML]{0070C0} 33.03} \\ \hline
\multicolumn{1}{|c|}{\textbf{NODE-ImgNet}}                     & \multicolumn{1}{c|}{{\color[HTML]{0070C0} 32.68}} & \multicolumn{1}{c|}{{\color[HTML]{C00000} 35.44}} & \multicolumn{1}{c|}{{\color[HTML]{C00000} 33.48}} & \multicolumn{1}{c|}{{\color[HTML]{C00000} 32.35}} & \multicolumn{1}{c|}{{\color[HTML]{0070C0} 33.29}} & \multicolumn{1}{c|}{{\color[HTML]{C00000} 31.90}} & \multicolumn{1}{c|}{{\color[HTML]{0070C0} 31.97}} & \multicolumn{1}{c|}{{\color[HTML]{C00000} 34.76}} & \multicolumn{1}{c|}{{\color[HTML]{0070C0} 32.87}} & \multicolumn{1}{c|}{{\color[HTML]{0070C0} 32.53}} & \multicolumn{1}{c|}{{\color[HTML]{C00000} 32.52}} & \multicolumn{1}{c|}{{\color[HTML]{C00000} 32.65}} & {\color[HTML]{C00000} 33.04} \\ \hline
\multicolumn{1}{|c|}{\textbf{NODE-ImgNet-B}}                   & \multicolumn{1}{c|}{32.55}                        & \multicolumn{1}{c|}{35.28}                        & \multicolumn{1}{c|}{33.39}                        & \multicolumn{1}{c|}{32.22}                        & \multicolumn{1}{c|}{33.10}                        & \multicolumn{1}{c|}{31.78}                        & \multicolumn{1}{c|}{31.88}                        & \multicolumn{1}{c|}{34.70}                        & \multicolumn{1}{c|}{32.35}                        & \multicolumn{1}{c|}{32.44}                        & \multicolumn{1}{c|}{32.44}                        & \multicolumn{1}{c|}{32.58}                        & 32.89                        \\ \hline
\multicolumn{14}{|c|}{\textbf{Noise Level $\sigma=25$}}                                                                                                                                                                                                                                                                                                                                                                                                                                                                                                                                                                                                                                                                                              \\ \hline
\multicolumn{1}{|c|}{\textbf{DnCNN \cite{zhang2017beyond}}}     & \multicolumn{1}{c|}{30.18}                        & \multicolumn{1}{c|}{33.06}                        & \multicolumn{1}{c|}{30.87}                        & \multicolumn{1}{c|}{29.41}                        & \multicolumn{1}{c|}{30.28}                        & \multicolumn{1}{c|}{29.13}                        & \multicolumn{1}{c|}{29.43}                        & \multicolumn{1}{c|}{32.44}                        & \multicolumn{1}{c|}{30.00}                        & \multicolumn{1}{c|}{30.21}                        & \multicolumn{1}{c|}{30.10}                        & \multicolumn{1}{c|}{30.12}                        & 30.43                        \\ \hline
\multicolumn{1}{|c|}{\textbf{FFDNet \cite{zhang2018ffdnet}}}    & \multicolumn{1}{c|}{30.10}                        & \multicolumn{1}{c|}{33.28}                        & \multicolumn{1}{c|}{30.93}                        & \multicolumn{1}{c|}{29.32}                        & \multicolumn{1}{c|}{30.08}                        & \multicolumn{1}{c|}{29.04}                        & \multicolumn{1}{c|}{29.44}                        & \multicolumn{1}{c|}{32.57}                        & \multicolumn{1}{c|}{30.01}                        & \multicolumn{1}{c|}{30.25}                        & \multicolumn{1}{c|}{30.11}                        & \multicolumn{1}{c|}{30.20}                        & 30.44                        \\ \hline
\multicolumn{1}{|c|}{\textbf{ADNet \cite{tian2020attention}}}   & \multicolumn{1}{c|}{30.34}                        & \multicolumn{1}{c|}{33.41}                        & \multicolumn{1}{c|}{31.14}                        & \multicolumn{1}{c|}{29.41}                        & \multicolumn{1}{c|}{30.39}                        & \multicolumn{1}{c|}{29.17}                        & \multicolumn{1}{c|}{29.49}                        & \multicolumn{1}{c|}{32.61}                        & \multicolumn{1}{c|}{30.25}                        & \multicolumn{1}{c|}{30.37}                        & \multicolumn{1}{c|}{30.08}                        & \multicolumn{1}{c|}{30.24}                        & 30.58                        \\ \hline
\multicolumn{1}{|c|}{\textbf{DudeNet \cite{tian2021designing}}} & \multicolumn{1}{c|}{30.23}                        & \multicolumn{1}{c|}{33.24}                        & \multicolumn{1}{c|}{30.98}                        & \multicolumn{1}{c|}{29.53}                        & \multicolumn{1}{c|}{30.44}                        & \multicolumn{1}{c|}{29.14}                        & \multicolumn{1}{c|}{29.48}                        & \multicolumn{1}{c|}{32.52}                        & \multicolumn{1}{c|}{30.15}                        & \multicolumn{1}{c|}{30.24}                        & \multicolumn{1}{c|}{30.08}                        & \multicolumn{1}{c|}{30.15}                        & 30.52                        \\ \hline
\multicolumn{1}{|c|}{\textbf{MWDCNN   \cite{tian2023multi}}}   & \multicolumn{1}{c|}{30.19}                        & \multicolumn{1}{c|}{33.33}                        & \multicolumn{1}{c|}{30.85}                        & \multicolumn{1}{c|}{{\color[HTML]{0070C0} 29.66}} & \multicolumn{1}{c|}{{\color[HTML]{0070C0} 30.55}} & \multicolumn{1}{c|}{29.16}                        & \multicolumn{1}{c|}{29.48}                        & \multicolumn{1}{c|}{32.67}                        & \multicolumn{1}{c|}{30.21}                        & \multicolumn{1}{c|}{30.28}                        & \multicolumn{1}{c|}{30.10}                        & \multicolumn{1}{c|}{30.13}                        & 30.55                        \\ \hline
\multicolumn{1}{|c|}{\textbf{BRDNet \cite{tian2020image}}}     & \multicolumn{1}{c|}{{\color[HTML]{C00000} 31.39}} & \multicolumn{1}{c|}{33.41}                        & \multicolumn{1}{c|}{31.04}                        & \multicolumn{1}{c|}{29.46}                        & \multicolumn{1}{c|}{30.50}                        & \multicolumn{1}{c|}{29.20}                        & \multicolumn{1}{c|}{{\color[HTML]{C00000} 29.55}} & \multicolumn{1}{c|}{32.65}                        & \multicolumn{1}{c|}{{\color[HTML]{0070C0} 30.34}} & \multicolumn{1}{c|}{{\color[HTML]{0070C0} 30.33}} & \multicolumn{1}{c|}{30.14}                        & \multicolumn{1}{c|}{30.28}                        & {\color[HTML]{0070C0} 30.61} \\ \hline
\multicolumn{1}{|c|}{\textbf{NODE-ImgNet}}                     & \multicolumn{1}{c|}{{\color[HTML]{0070C0} 30.33}} & \multicolumn{1}{c|}{{\color[HTML]{C00000} 33.68}} & \multicolumn{1}{c|}{{\color[HTML]{C00000} 31.12}} & \multicolumn{1}{c|}{{\color[HTML]{C00000} 29.82}} & \multicolumn{1}{c|}{{\color[HTML]{C00000} 30.57}} & \multicolumn{1}{c|}{{\color[HTML]{C00000} 29.31}} & \multicolumn{1}{c|}{{\color[HTML]{0070C0} 29.55}} & \multicolumn{1}{c|}{{\color[HTML]{C00000} 32.80}} & \multicolumn{1}{c|}{{\color[HTML]{C00000} 30.56}} & \multicolumn{1}{c|}{{\color[HTML]{C00000} 30.41}} & \multicolumn{1}{c|}{{\color[HTML]{C00000} 30.21}} & \multicolumn{1}{c|}{{\color[HTML]{C00000} 30.43}} & {\color[HTML]{C00000} 30.73} \\ \hline
\multicolumn{1}{|c|}{\textbf{NODE-ImgNet-B}}                   & \multicolumn{1}{c|}{30.22}                        & \multicolumn{1}{c|}{{\color[HTML]{0070C0} 33.61}} & \multicolumn{1}{c|}{{\color[HTML]{2F75B5} 31.07}} & \multicolumn{1}{c|}{29.64}                        & \multicolumn{1}{c|}{30.51}                        & \multicolumn{1}{c|}{{\color[HTML]{2F75B5} 29.23}} & \multicolumn{1}{c|}{29.51}                        & \multicolumn{1}{c|}{{\color[HTML]{2F75B5} 32.74}} & \multicolumn{1}{c|}{29.90}                        & \multicolumn{1}{c|}{{\color[HTML]{2F75B5} 30.33}} & \multicolumn{1}{c|}{{\color[HTML]{2F75B5} 30.19}} & \multicolumn{1}{c|}{{\color[HTML]{2F75B5} 30.35}} & {\color[HTML]{2F75B5} 30.61} \\ \hline
\multicolumn{14}{|c|}{\textbf{Noise Level $\sigma= 50$}}                                                                                                                                                                                                                                                                                                                                                                                                                                                                                                                                                                                                                                                                                              \\ \hline
\multicolumn{1}{|c|}{\textbf{DnCNN \cite{zhang2017beyond}}}     & \multicolumn{1}{c|}{27.03}                        & \multicolumn{1}{c|}{30.00}                        & \multicolumn{1}{c|}{27.32}                        & \multicolumn{1}{c|}{25.70}                        & \multicolumn{1}{c|}{26.78}                        & \multicolumn{1}{c|}{25.87}                        & \multicolumn{1}{c|}{26.48}                        & \multicolumn{1}{c|}{29.39}                        & \multicolumn{1}{c|}{26.22}                        & \multicolumn{1}{c|}{27.20}                        & \multicolumn{1}{c|}{27.24}                        & \multicolumn{1}{c|}{26.90}                        & 27.18                        \\ \hline
\multicolumn{1}{|c|}{\textbf{FFDNet \cite{zhang2018ffdnet}}}    & \multicolumn{1}{c|}{27.05}                        & \multicolumn{1}{c|}{30.37}                        & \multicolumn{1}{c|}{27.54}                        & \multicolumn{1}{c|}{25.75}                        & \multicolumn{1}{c|}{26.81}                        & \multicolumn{1}{c|}{25.89}                        & \multicolumn{1}{c|}{26.57}                        & \multicolumn{1}{c|}{29.66}                        & \multicolumn{1}{c|}{26.45}                        & \multicolumn{1}{c|}{27.33}                        & \multicolumn{1}{c|}{27.29}                        & \multicolumn{1}{c|}{27.08}                        & 27.32                        \\ \hline
\multicolumn{1}{|c|}{\textbf{ADNet \cite{tian2020attention}}}   & \multicolumn{1}{c|}{27.31}                        & \multicolumn{1}{c|}{30.59}                        & \multicolumn{1}{c|}{27.69}                        & \multicolumn{1}{c|}{25.70}                        & \multicolumn{1}{c|}{26.90}                        & \multicolumn{1}{c|}{25.88}                        & \multicolumn{1}{c|}{26.56}                        & \multicolumn{1}{c|}{29.59}                        & \multicolumn{1}{c|}{26.64}                        & \multicolumn{1}{c|}{27.35}                        & \multicolumn{1}{c|}{27.17}                        & \multicolumn{1}{c|}{27.07}                        & 27.37                        \\ \hline
\multicolumn{1}{|c|}{\textbf{DudeNet \cite{tian2021designing}}} & \multicolumn{1}{c|}{27.22}                        & \multicolumn{1}{c|}{30.27}                        & \multicolumn{1}{c|}{27.51}                        & \multicolumn{1}{c|}{25.88}                        & \multicolumn{1}{c|}{26.93}                        & \multicolumn{1}{c|}{25.88}                        & \multicolumn{1}{c|}{26.50}                        & \multicolumn{1}{c|}{29.45}                        & \multicolumn{1}{c|}{26.49}                        & \multicolumn{1}{c|}{27.26}                        & \multicolumn{1}{c|}{27.19}                        & \multicolumn{1}{c|}{26.97}                        & 27.30                        \\ \hline
\multicolumn{1}{|c|}{\textbf{MWDCNN \cite{tian2023multi}}}     & \multicolumn{1}{c|}{26.99}                        & \multicolumn{1}{c|}{30.58}                        & \multicolumn{1}{c|}{27.34}                        & \multicolumn{1}{c|}{25.85}                        & \multicolumn{1}{c|}{27.02}                        & \multicolumn{1}{c|}{25.93}                        & \multicolumn{1}{c|}{26.48}                        & \multicolumn{1}{c|}{29.63}                        & \multicolumn{1}{c|}{26.60}                        & \multicolumn{1}{c|}{27.23}                        & \multicolumn{1}{c|}{27.27}                        & \multicolumn{1}{c|}{27.11}                        & 27.34                        \\ \hline
\multicolumn{1}{|c|}{\textbf{BRDNet   \cite{tian2020image}}}   & \multicolumn{1}{c|}{{\color[HTML]{C00000} 27.44}} & \multicolumn{1}{c|}{30.53}                        & \multicolumn{1}{c|}{27.67}                        & \multicolumn{1}{c|}{25.77}                        & \multicolumn{1}{c|}{26.97}                        & \multicolumn{1}{c|}{25.93}                        & \multicolumn{1}{c|}{26.66}                        & \multicolumn{1}{c|}{29.73}                        & \multicolumn{1}{c|}{26.85}                        & \multicolumn{1}{c|}{27.38}                        & \multicolumn{1}{c|}{27.27}                        & \multicolumn{1}{c|}{27.17}                        & 27.45                        \\ \hline
\multicolumn{1}{|c|}{\textbf{NODE-ImgNet}}                     & \multicolumn{1}{c|}{{\color[HTML]{0070C0} 27.32}} & \multicolumn{1}{c|}{{\color[HTML]{C00000} 31.10}} & \multicolumn{1}{c|}{{\color[HTML]{C00000} 27.73}} & \multicolumn{1}{c|}{{\color[HTML]{C00000} 26.05}} & \multicolumn{1}{c|}{{\color[HTML]{C00000} 27.18}} & \multicolumn{1}{c|}{{\color[HTML]{C00000} 26.03}} & \multicolumn{1}{c|}{{\color[HTML]{2F75B5} 26.67}} & \multicolumn{1}{c|}{{\color[HTML]{C00000} 29.90}} & \multicolumn{1}{c|}{{\color[HTML]{C00000} 27.32}} & \multicolumn{1}{c|}{{\color[HTML]{C00000} 27.46}} & \multicolumn{1}{c|}{{\color[HTML]{C00000} 27.41}} & \multicolumn{1}{c|}{{\color[HTML]{C00000} 27.37}} & {\color[HTML]{C00000} 27.63} \\ \hline
\multicolumn{1}{|c|}{\textbf{NODE-ImgNet-B}}                   & \multicolumn{1}{c|}{27.30}                        & \multicolumn{1}{c|}{{\color[HTML]{2F75B5} 31.03}} & \multicolumn{1}{c|}{{\color[HTML]{2F75B5} 27.68}} & \multicolumn{1}{c|}{{\color[HTML]{2F75B5} 25.98}} & \multicolumn{1}{c|}{{\color[HTML]{2F75B5} 27.17}} & \multicolumn{1}{c|}{{\color[HTML]{2F75B5} 26.00}} & \multicolumn{1}{c|}{{\color[HTML]{C00000} 26.68}} & \multicolumn{1}{c|}{{\color[HTML]{2F75B5} 29.88}} & \multicolumn{1}{c|}{{\color[HTML]{2F75B5} 27.08}} & \multicolumn{1}{c|}{{\color[HTML]{2F75B5} 27.41}} & \multicolumn{1}{c|}{{\color[HTML]{2F75B5} 27.36}} & \multicolumn{1}{c|}{{\color[HTML]{2F75B5} 27.33}} & {\color[HTML]{2F75B5} 27.57} \\ \hline
\end{tabular}%
}
\caption{
PSNR (dB) results of various methods for Gaussian gray image denoising on the Set12 Dataset at various noise levels.}
\label{tab:gray_set12}
\end{table}

In summary, we conclude that NODE-ImgNet provides a highly competitive model for image denoising on gray images with Gaussian noise. Moreover, as noise level increases, we observe that the NODE-ImgNet achieves more significant improvement than baseline models.


\subsubsection{NODE-ImgNet for color Gaussian image denoising}
\label{sec:color Gaussian}
{For color Gaussian image denoising, 
we also train multiple NODE-ImgNet models in two modes. In the first mode, we train $5$ separate NODE-ImgNet models with fixed noise levels $\sigma = 15, 25, 35, 50, 75$, respectively. 
In the second mode, we employ a similar blind denoising approach as in \cref{sec:gray Gaussian},
i.e., the noise level $\sigma$ for each image patch is uniformly distributed between $0$ and $55$ during the training process. 
This model is also referred to as NODE-ImgNet-B.} 

We assess the denoising performance of our models with other benchmark models on datasets CBSD68, Kodak24, and McMaster. The results are presented in \Cref{color_all_test_table}. It is obvious that NODE-ImgNet demonstrates uniform superior effectiveness for color Gaussian image denoising in comparison to other baseline models.
In addition, it is worth noting that our NODE-ImgNet-B model also outperforms other baseline models at $\sigma = 25,35 ,50$ for all test datasets, demonstrating superior blind noise reduction capabilities. 


\begin{table}[]
\resizebox{\columnwidth}{!}{%
\begin{tabular}{|c|c|c|c|c|c|c|}
\hline
Datasets                   & {\color[HTML]{2E2E2E} \textbf{Methods}}                & {\color[HTML]{2E2E2E} \textbf{$\sigma = 15$}} & {\color[HTML]{2E2E2E} \textbf{$\sigma = 25$}} & {\color[HTML]{2E2E2E} \textbf{$\sigma = 35$}} & {\color[HTML]{2E2E2E} \textbf{$\sigma = 50$}} & {\color[HTML]{2E2E2E} \textbf{$\sigma = 75$}} \\ \hline
                           & DnCNN \cite{zhang2017beyond}                            & 33.98                                         & 31.31                                         & 29.65                                         & 28.01                                         & -                                             \\ \cline{2-7} 
                           & DC-B-R \cite{wang2017dilated}              & 31.68                                         & -                                             & 29.18                                         & 26.21                                         & -                                             \\ \cline{2-7} 
                           & FFDNet \cite{zhang2018ffdnet}                           & 33.80                                         & 31.18                                         & 29.57                                         & 27.96                                         & 26.24                                         \\ \cline{2-7} 
                           & GradNet \cite{liu2020gradnet}                           & 34.07                                         & 31.39                                         & -                                             & 28.12                                         & -                                             \\ \cline{2-7} 
                           & ADNet \cite{tian2020attention}                          & 33.99                                         & 31.31                                         & 29.66                                         & 28.04                                         & 26.33                                         \\ \cline{2-7} 
                           & DudeNet \cite{tian2021designing}                        & 34.01                                         & 31.34                                         & 29.71                                         & 28.09                                         & 26.40                                         \\ \cline{2-7} 
                           & BRDNet \cite{tian2020image}                      & {\color[HTML]{00B0F0} 34.10}                  & 31.43                                         & 29.77                                         & 28.16                                         & {\color[HTML]{00B0F0} 26.43}                  \\ \cline{2-7} 
                           & NODE-ImgNet                                            & {\color[HTML]{FF0000} 34.19}                  & {\color[HTML]{FF0000} 31.49}                  & {\color[HTML]{FF0000} 29.88}                  & {\color[HTML]{FF0000} 28.29}                  & {\color[HTML]{FF0000} 26.60}                  \\ \cline{2-7} 
\multirow{-9}{*}{CBSD68}   & NODE-ImgNet-B                                          & 34.05                                         & {\color[HTML]{00B0F0} 31.45}                  & {\color[HTML]{00B0F0} 29.86}                  & {\color[HTML]{00B0F0} 28.24}                  & {\color[HTML]{2E2E2E} -}                      \\ \hline
                           & DnCNN \cite{zhang2017beyond}                            & {\color[HTML]{2E2E2E} 34.73}                  & {\color[HTML]{2E2E2E} 32.23}                  & {\color[HTML]{2E2E2E} 30.64}                  & {\color[HTML]{2E2E2E} 29.02}                  & {\color[HTML]{2E2E2E} -}                      \\ \cline{2-7} 
                           & FFDNet \cite{zhang2018ffdnet}                           & {\color[HTML]{2E2E2E} 34.55}                  & {\color[HTML]{2E2E2E} 32.11}                  & {\color[HTML]{2E2E2E} 30.56}                  & {\color[HTML]{2E2E2E} 28.99}                  & {\color[HTML]{2E2E2E} 27.25}                  \\ \cline{2-7} 
                           & GradNet \cite{liu2020gradnet}                           & {\color[HTML]{2E2E2E} 34.85}                  & {\color[HTML]{2E2E2E} 32.35}                  & {\color[HTML]{2E2E2E} -}                      & {\color[HTML]{2E2E2E} 29.23}                  & {\color[HTML]{2E2E2E} -}                      \\ \cline{2-7} 
                           & ADNet \cite{tian2020attention}                          & 34.76                                         & 32.26                                         & 30.68                                         & 29.1                                          & 27.4                                          \\ \cline{2-7} 
                           & {\color[HTML]{2E2E2E} DudeNet \cite{tian2021designing}} & 34.81                                         & 32.26                                         & 30.69                                         & 29.1                                          & 27.39                                         \\ \cline{2-7} 
                           & BRDNet \cite{tian2020image}                      & {\color[HTML]{00B0F0} 34.88}                  & 32.41                                         & 30.8                                          & 29.22                                         & {\color[HTML]{00B0F0} 27.49}                  \\ \cline{2-7} 
                           & NODE-ImgNet                                            & {\color[HTML]{FF0000} 34.96}                  & {\color[HTML]{FF0000} 32.51}                  & {\color[HTML]{FF0000} 31.00}                  & {\color[HTML]{FF0000} 29.47}                  & {\color[HTML]{FF0000} 27.78}                  \\ \cline{2-7} 
\multirow{-8}{*}{Kodak24}  & NODE-ImgNet-B                                          & 34.87                                         & {\color[HTML]{00B0F0} 32.47}                  & {\color[HTML]{00B0F0} 30.96}                  & {\color[HTML]{00B0F0} 29.38}                  & {\color[HTML]{2E2E2E} -}                      \\ \hline
                           & DnCNN \cite{zhang2017beyond}                            & {\color[HTML]{2E2E2E} 34.80}                  & {\color[HTML]{2E2E2E} 32.47}                  & {\color[HTML]{2E2E2E} 30.91}                  & {\color[HTML]{2E2E2E} 29.21}                  & {\color[HTML]{2E2E2E} -}                      \\ \cline{2-7} 
                           & FFDNet \cite{zhang2018ffdnet}                           & {\color[HTML]{2E2E2E} 34.47}                  & {\color[HTML]{2E2E2E} 32.25}                  & {\color[HTML]{2E2E2E} 30.76}                  & {\color[HTML]{2E2E2E} 29.21}                  & {\color[HTML]{2E2E2E} 27.29}                  \\ \cline{2-7} 
                           & GradNet \cite{liu2020gradnet}                           & {\color[HTML]{2E2E2E} 34.81}                  & {\color[HTML]{2E2E2E} 32.45}                  & {\color[HTML]{2E2E2E} -}                      & {\color[HTML]{2E2E2E} 29.39}                  & {\color[HTML]{2E2E2E} -}                      \\ \cline{2-7} 
                           & ADNet \cite{tian2020attention}                          & 34.93                                         & 32.56                                         & 31                                            & 29.36                                         & 27.53                                         \\ \cline{2-7} 
                           & BRDNet \cite{tian2020image}                      & {\color[HTML]{00B0F0} 35.08}                  & 32.75                                         & 31.15                                         & 29.52                                         & {\color[HTML]{00B0F0} 27.72}                  \\ \cline{2-7} 
                           & NODE-ImgNet                                            & {\color[HTML]{FF0000} 35.11}                  & {\color[HTML]{FF0000} 32.83}                  & {\color[HTML]{FF0000} 31.35}                  & {\color[HTML]{FF0000} 29.79}                  & {\color[HTML]{FF0000} 27.97}                  \\ \cline{2-7} 
\multirow{-7}{*}{McMaster} & NODE-ImgNet-B                                          & 35.01                                         & {\color[HTML]{00B0F0} 32.78}                  & {\color[HTML]{00B0F0} 31.31}                  & {\color[HTML]{00B0F0} 29.72}                  & {\color[HTML]{2E2E2E} -}                      \\ \hline
\end{tabular}%
}
\caption{Average PSNR (dB) results of various methods for Gaussian color image denoising at various noise levels on the CBSD68, Kodak24, and McMaster datasets.}
\label{color_all_test_table}
\end{table}

\begin{figure}[hbt!]
    \centering
    \begin{subfigure}{0.32\textwidth}
        \includegraphics[width=\textwidth]{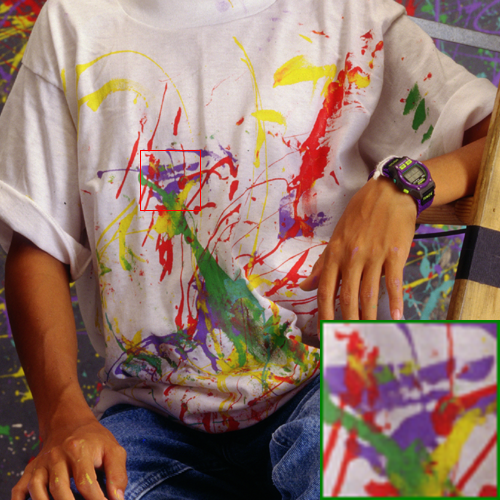}
        \caption{Clean image}
        \label{fig:McMaster image1}
    \end{subfigure}
    \hspace{0em} 
    \begin{subfigure}{0.32\textwidth}
        \includegraphics[width=\textwidth]{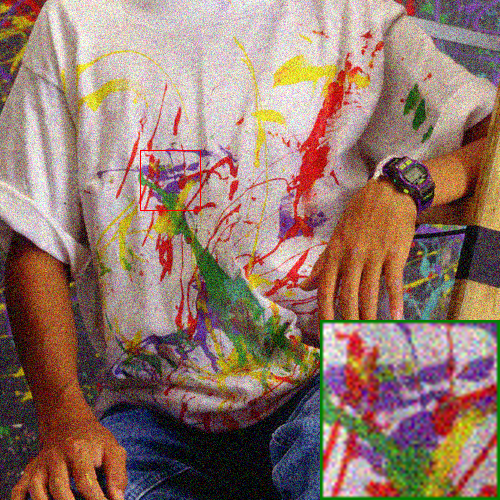}
        \caption{Noisy image ($23.89$dB)}
        \label{fig:McMaster image2}
    \end{subfigure}
    \hspace{0em} 
    \begin{subfigure}{0.32\textwidth}
        \includegraphics[width=\textwidth]{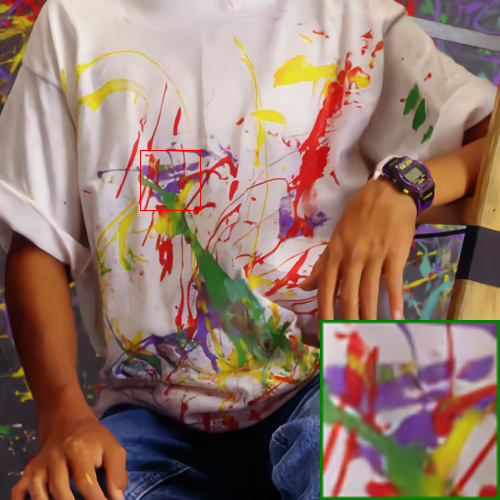}
        \caption{FFDNet ($33.28$dB)}
        \label{fig:McMaster image3}
    \end{subfigure}

    \vspace{0.3em}
    \begin{subfigure}{0.32\textwidth}
        \includegraphics[width=\textwidth]{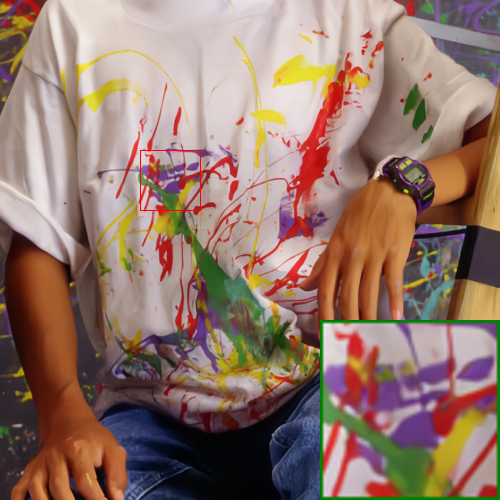}
        \caption{BrdNet ($33.54$dB)}
        \label{fig:McMaster image4}
    \end{subfigure}
    \hspace{0em} 
    \begin{subfigure}{0.32\textwidth}
        \includegraphics[width=\textwidth]{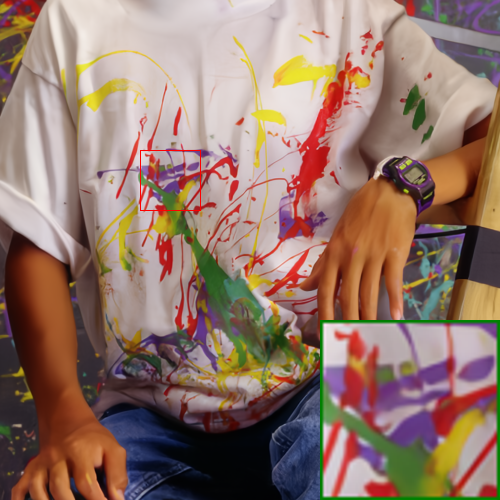}
        \caption{\color{red}NODE-ImgNet-B ($33.59$dB)}
        \label{fig:McMaster image5}
    \end{subfigure}
    \hspace{0em} 
    \begin{subfigure}{0.32\textwidth}
        \includegraphics[width=\textwidth]{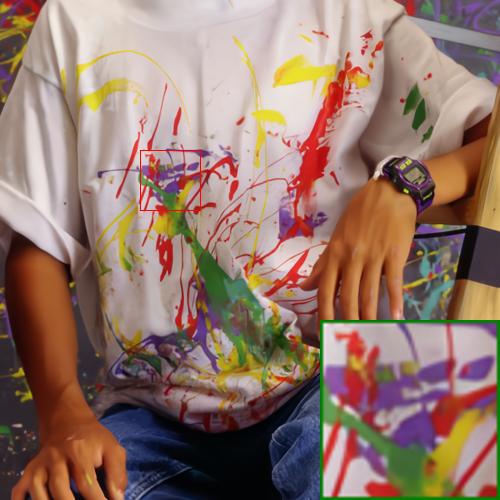}
        \caption{NODE-ImgNet ($33.65$dB)}
        \label{fig:McMaster image6}
    \end{subfigure}
    
\caption{Denoising results for a Gaussian color noisy image from the McMaster dataset with $\sigma = 25$.}
\label{pic_diff_25}
\end{figure}

\begin{figure}[hbt!]
    \centering
    \begin{subfigure}{0.32\textwidth}
        \includegraphics[width=\textwidth]{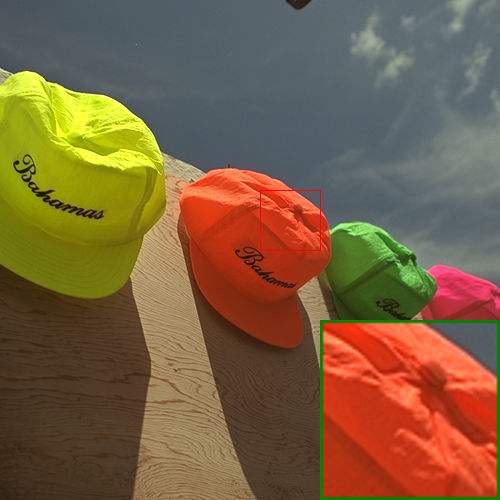}
        \caption{Clean image}
        \label{fig:Kodak24 image1}
    \end{subfigure}
    \hspace{0em} 
    \begin{subfigure}{0.32\textwidth}
    \includegraphics[width=\textwidth]{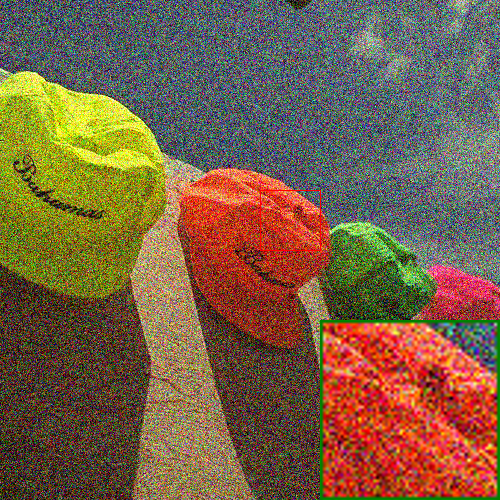}
        \caption{Noisy image ($18.16$dB)}
        \label{fig:Kodak24 image2}
    \end{subfigure}
    \hspace{0em} 
    \begin{subfigure}{0.32\textwidth}
    \includegraphics[width=\textwidth]{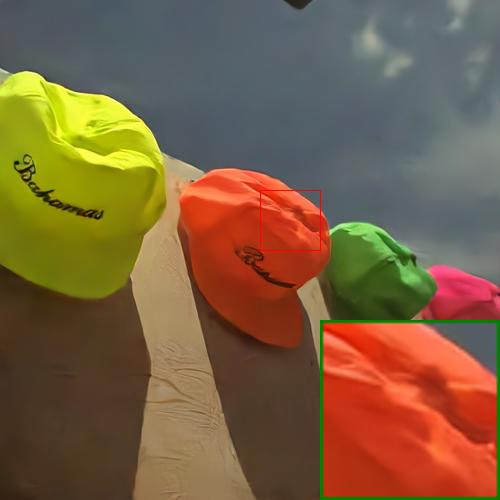}
        \caption{FFDNet ($32.85$dB)}
        \label{fig:Kodak24 image3}
    \end{subfigure}
    \vspace{0.3em}
    \begin{subfigure}{0.32\textwidth}
    \includegraphics[width=\textwidth]{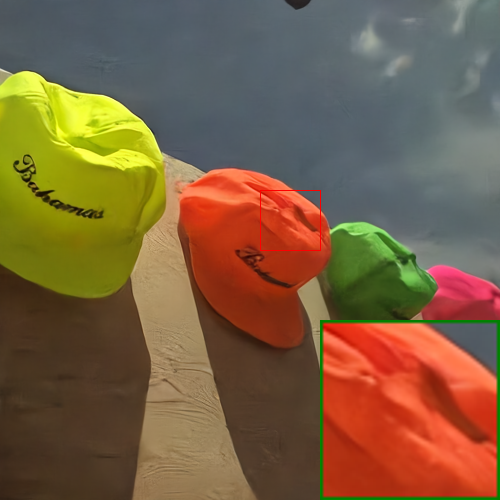}
        \caption{BRDNet ($33.05$dB)}
        \label{fig:Kodak24 image4}
    \end{subfigure}
    \hspace{0em} 
    \begin{subfigure}{0.32\textwidth}
    \includegraphics[width=\textwidth]{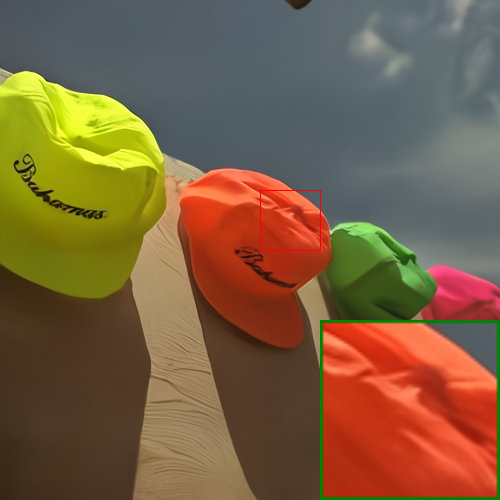}
        \caption{NODE-ImgNet-B ($33.35$dB)}
        \label{fig:Kodak24 image5}
    \end{subfigure}
    \hspace{0em} 
    \begin{subfigure}{0.32\textwidth}
    \includegraphics[width=\textwidth]{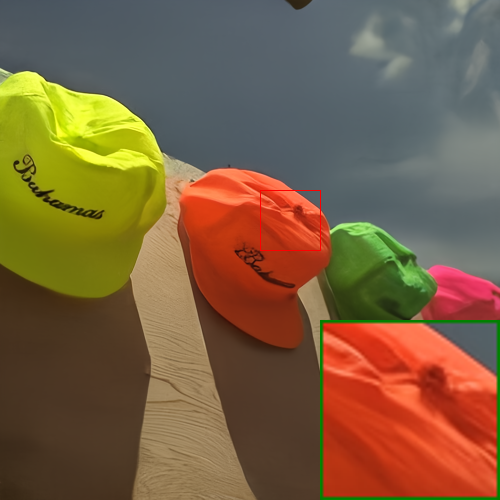}
        \caption{NODE-ImgNet ($33.50$dB)}
        \label{fig:Kodak24 image6}
    \end{subfigure}
\caption{Denoising results for a Gaussian color noisy image from the Kodak24 dataset with $\sigma = 50$.}
\label{pic_diff_50}
\end{figure}

\begin{figure}[hbt!]
    \centering
    \begin{subfigure}{0.32\textwidth}
        \includegraphics[width=\textwidth]{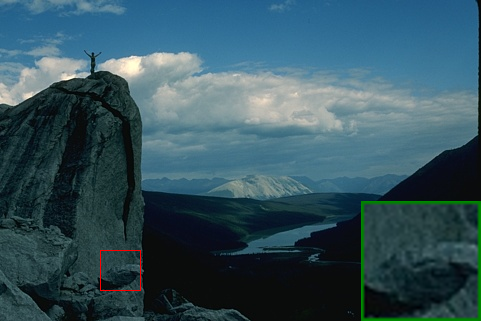}
        \caption{Clean image}
        \label{fig:CBSD68 image1}
    \end{subfigure}
    \hspace{0em} 
    \begin{subfigure}{0.32\textwidth}
        \includegraphics[width=\textwidth]{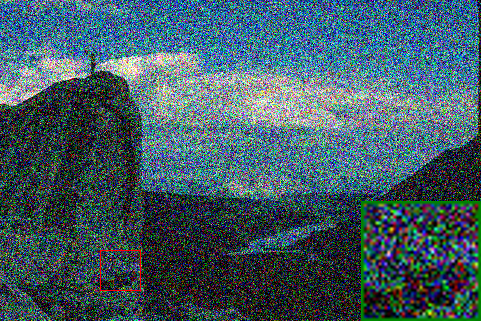}
        \caption{Noisy image ($15.48$dB)}
        \label{fig:CBSD68 image3}
    \end{subfigure}

    \vspace{0.3em}
    \begin{subfigure}{0.32\textwidth}
        \includegraphics[width=\textwidth]{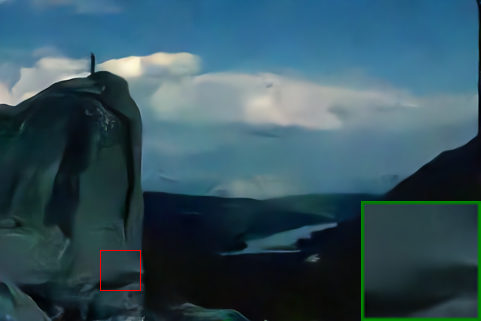}
        \caption{FFDNet ($30.68$dB)}
        \label{fig:CBSD68 image4}
    \end{subfigure}
    \hspace{0em} 
    \begin{subfigure}{0.32\textwidth}
        \includegraphics[width=\textwidth]{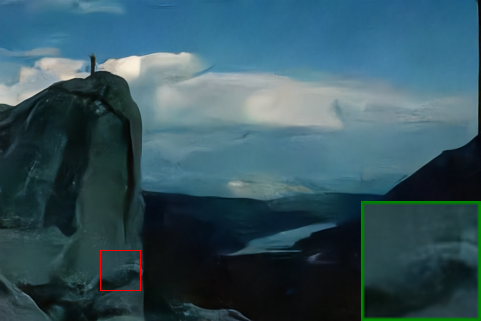}
        \caption{BRDNet ($30.94$dB)}
        \label{fig:CBSD68 image5}
    \end{subfigure}
    \hspace{0em} 
    \begin{subfigure}{0.32\textwidth}
        \includegraphics[width=\textwidth]{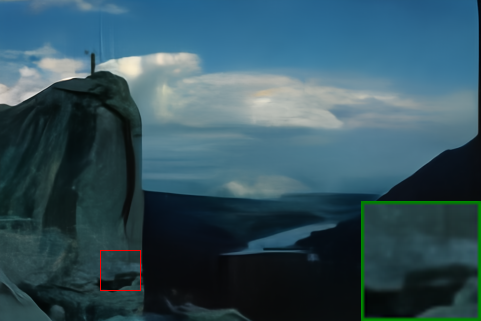}
        \caption{NODE-ImgNet ($31.12$dB)}
        \label{fig:CBSD68 image6}
    \end{subfigure}

\caption{Denoising results for a Gaussian color noisy image from the CBSD68 dataset with $\sigma = 75$.}
\label{fig:diff_75}
\end{figure}

Figures \ref{pic_diff_25},  \ref{pic_diff_50} and \ref{fig:diff_75} visualize  the resulting denoised color images from various denoising models at different Gaussian noise, i.e., $\sigma = 25$, $50$ and $75$. It is evident from each of the figures that the denoised image produced by NODE-ImgNet successfully restores the most texture and details  depicted in the clean image.


{In summary, NODE-ImgNet stands out as a competitive model for denoising color images affected by Gaussian noise. Once again, as the noise level escalates, NODE-ImgNet exhibits a considerably more substantial improvement in performance in contrast to the baseline models.}

\subsubsection{NODE-ImgNet for real image denoising}
\label{sec:real noise}
In this subsection, we compare the performance of NODE-ImgNet on real image denoising with other 
competitive baseline methods
 including DnCNN \cite{zhang2017beyond}, ADNet \cite{tian2020attention}, DudeNet \cite{tian2021designing}, MWDCNN \cite{tian2023multi} and BRDNet \cite{tian2020image}.
The trained model is tested on the cc \cite{nam2016holistic} dataset with $15$ noisy images.  
{The results presented in  \Cref{tab:real} show that NODE-ImgNet outperforms other models for the majority of images in the data set.
{In particular, when the ISO values are larger, indicating higher levels of noise, our model performs uniformly better with higher improvement.} 
{This along with former results indicates that our NODE-ImgNet has strong denoising capabilities for more complex noise types, such as real-world noise.}

{It is important to note that for real image denoising, the improvement of NODE-ImgNet over other baseline networks, e.g., BRDNet, is more significant than the synthetic Gaussian denoising tests.}
{More precisely, for the color Gaussian denoising experiments, the average improvement range in PSNR compared to BRDNet across each test set is between $0.03$ to $0.29$ dB. Whereas, for the real image denoising, the improvement is $0.60$ in dB which is more than $2$ folds higher.}

More precisely, for the color Gaussian denoising experiments, the improvement range in PSNR compared to BRDNet is between $0.03$ to $0.29$ in dB. Whereas, for the real image denoising, the improvement is 0.6  dB which is about $2$  times higher.



\begin{table}[h!]
\resizebox{\columnwidth}{!}{%
\begin{tabular}{|c|c|c|c|c|c|c|}
\hline
Camera   settings                       & DnCNN \cite{zhang2017beyond} & ADNet \cite{tian2020attention} & DudeNet \cite{tian2021designing} & MWDCNN \cite{tian2023multi}  & BRDNet \cite{tian2020image}  & NODE-ImgNet                  \\ \hline
                                        & {\color[HTML]{0070c0} 37.26} & 35.96                          & 36.66                            & 36.97                        & {\color[HTML]{c00000} 37.63} & 37.09                        \\ \cline{2-7} 
                                        & 34.13                        & 36.11                          & 36.70                            & 36.01                        & {\color[HTML]{c00000} 37.28} & {\color[HTML]{0070c0} 36.78} \\ \cline{2-7} 
\multirow{-3}{*}{Canon   5D ISO=3200}   & 34.09                        & 34.49                          & 35.03                            & 34.80                        & {\color[HTML]{c00000} 37.75} & {\color[HTML]{0070c0} 35.58} \\ \hline
                                        & 33.62                        & 33.94                          & 33.72                            & 33.91                        & {\color[HTML]{0070c0} 34.55} & {\color[HTML]{c00000} 34.93} \\ \cline{2-7} 
                                        & 34.48                        & 34.33                          & 34.70                            & 34.88                        & {\color[HTML]{0070c0} 35.99} & {\color[HTML]{c00000} 36.43} \\ \cline{2-7} 
\multirow{-3}{*}{Nikon   D600 ISO=3200} & 35.41                        & {\color[HTML]{0070c0} 38.87}   & 37.98                            & 37.02                        & 38.62                        & {\color[HTML]{c00000} 42.05} \\ \hline
                                        & 35.79                        & 37.61                          & 38.10                            & {\color[HTML]{0070c0} 37.93} & {\color[HTML]{c00000} 39.22} & 37.92                        \\ \cline{2-7} 
                                        & 36.08                        & 38.24                          & 39.15                            & 37.49                        & {\color[HTML]{0070c0} 39.67} & {\color[HTML]{c00000} 39.81} \\ \cline{2-7} 
\multirow{-3}{*}{Nikon   D800 ISO=1600} & 35.48                        & 36.89                          & 36.14                            & {\color[HTML]{0070c0} 38.44} & {\color[HTML]{c00000} 39.04} & 38.14                        \\ \hline
                                        & 34.08                        & 37.20                          & 36.96                            & 37.10                        & {\color[HTML]{0070c0} 38.28} & {\color[HTML]{c00000} 40.35} \\ \cline{2-7} 
                                        & 33.70                        & 35.67                          & 35.80                            & 36.72                        & {\color[HTML]{c00000} 37.18} & {\color[HTML]{0070c0} 36.76} \\ \cline{2-7} 
\multirow{-3}{*}{Nikon   D800 ISO=3200} & 33.31                        & 38.09                          & 37.49                            & 37.25                        & {\color[HTML]{0070c0} 38.85} & {\color[HTML]{c00000} 40.93} \\ \hline
                                        & 29.83                        & 32.24                          & 31.94                            & 32.24                        & {\color[HTML]{0070c0} 32.75} & {\color[HTML]{c00000} 34.79} \\ \cline{2-7} 
                                        & 30.55                        & 32.59                          & 32.51                            & 32.56                        & {\color[HTML]{0070c0} 33.24} & {\color[HTML]{c00000} 34.14} \\ \cline{2-7} 
\multirow{-3}{*}{Nikon   D800 ISO=6400} & 30.09                        & {\color[HTML]{0070c0} 33.14}   & 32.91                            & 32.76                        & 32.89                        & {\color[HTML]{c00000} 34.34} \\ \hline
Average                                 & 33.86                        & 35.69                          & 35.72                            & 35.74                        & {\color[HTML]{0070c0} 36.73} & {\color[HTML]{c00000} 37.33} \\ \hline
\end{tabular}%
}
\caption{PSNR (dB) Results of Various Methods on Real Noisy Images}
\label{tab:real}
\end{table}


\subsection{ 
{
Efficiency with Small-Scale Partial Datasets}}\label{sec:ComponentAnalysis}
\label{advantage1}
As aforementioned, the inherent flexibility in NODE-ImgNet enables its efficacy on smaller datasets, which is critical for tasks with limited access to training data.
To better support the statement,
 we train our NODE-ImgNet and BRDNet models based on each dataset that comprises  $50\%, 25\%, 10\%,$ and $2\%$ of the Waterloo Exploration Database with random sampling, respectively.
 The resulting number of training images for the four experiments is, therefore, $2372, 1186, 474$, and $95$, respectively. {Each selected image were also cropped into $300$ patches of size $60\times60$, following the previous way.}  The models are trained with fixed noise level $\sigma=50$.
 

The test results on the McMaster, Kodak24, and CBSD68 dataset are presented in \Cref{tab:small_dataset}. 
{We observe that NODE-ImgNet is less negatively affected when decreasing the amount of training data. To better visualize the data, we also organize them in \Cref{pic:small_data} for the McMaster test dataset.}
{
With linear regression,
the decrease rates for NODE-ImgNet and BRDNet is $-0.168$ and $-0.246$ in dB/percent, respectively.
Therefore, \Cref{pic:small_data} further confirms that our model can achieve better performance on {small-scale partial} datasets compared to BRDNet and, therefore, can effectively handle tasks with limited training data.}

\begin{figure}[h!]
\centering
\includegraphics[width=0.6\textwidth]{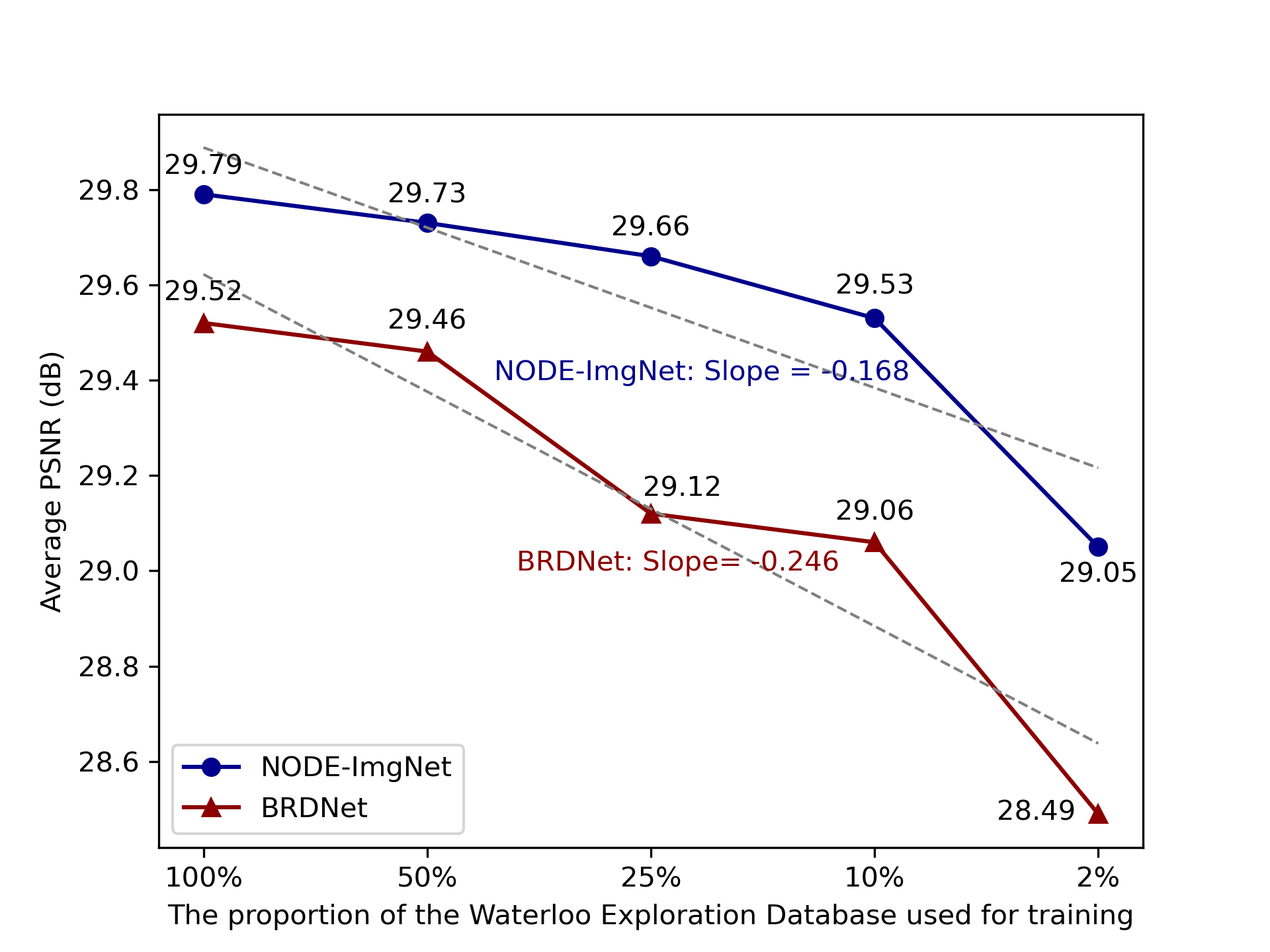}
\caption{Performance of NODE-ImgNet and BRDNet with decreasing size of training Dataset. Shown results are obtained through the McMaster test set.
}
\label{pic:small_data}
\end{figure}

\begin{table}[]
\resizebox{\columnwidth}{!}{%
\begin{tabular}{|c|c|c|c|cc|cc|}
\hline
                                    & \textbf{Training Data   Ratio} & \textbf{NODE-ImgNet}         & \textbf{BRDNet}              & \multicolumn{2}{c|}{\textbf{Relative   Decrease Value (dB)}}                     & \multicolumn{2}{c|}{\textbf{Relative   Decrease Percentage}}                         \\ \hline
                                    & \textbf{100\%}                 & {\color[HTML]{C00000} 29.79} & {\color[HTML]{0070C0} 29.52} & \multicolumn{1}{c|}{\textbf{NODE-ImgNet}}         & \textbf{BRDNet}              & \multicolumn{1}{c|}{\textbf{NODE-ImgNet}}           & \textbf{BRDNet}                \\ \cline{2-8} 
                                    & \textbf{50\%}                  & {\color[HTML]{C00000} 29.73} & {\color[HTML]{0070C0} 29.46} & \multicolumn{1}{c|}{{\color[HTML]{C00000} -0.06}} & {\color[HTML]{0070C0} -0.06} & \multicolumn{1}{c|}{{\color[HTML]{C00000} -0.19\%}} & {\color[HTML]{0070C0} -0.19\%} \\ \cline{2-8} 
                                    & \textbf{25\%}                  & {\color[HTML]{C00000} 29.66} & {\color[HTML]{0070C0} 29.12} & \multicolumn{1}{c|}{{\color[HTML]{C00000} -0.13}} & {\color[HTML]{0070C0} -0.40} & \multicolumn{1}{c|}{{\color[HTML]{C00000} -0.45\%}} & {\color[HTML]{0070C0} -1.37\%} \\ \cline{2-8} 
                                    & \textbf{10\%}                  & {\color[HTML]{C00000} 29.53} & {\color[HTML]{0070C0} 29.06} & \multicolumn{1}{c|}{{\color[HTML]{C00000} -0.26}} & {\color[HTML]{0070C0} -0.47} & \multicolumn{1}{c|}{{\color[HTML]{C00000} -0.87\%}} & {\color[HTML]{0070C0} -1.60\%} \\ \cline{2-8} 
                                    & \textbf{5\%}                   & {\color[HTML]{C00000} 29.29} & {\color[HTML]{0070C0} 29.02} & \multicolumn{1}{c|}{{\color[HTML]{C00000} -0.50}} & {\color[HTML]{0070C0} -0.50} & \multicolumn{1}{c|}{{\color[HTML]{C00000} -1.72\%}} & {\color[HTML]{0070C0} -1.74\%} \\ \cline{2-8} 
\multirow{-6}{*}{\textbf{McMaster}} & \textbf{2\%}                   & {\color[HTML]{C00000} 29.05} & {\color[HTML]{0070C0} 28.49} & \multicolumn{1}{c|}{{\color[HTML]{C00000} -0.74}} & {\color[HTML]{0070C0} -1.03} & \multicolumn{1}{c|}{{\color[HTML]{C00000} -2.56\%}} & {\color[HTML]{0070C0} -3.61\%} \\ \hline
                                    & \textbf{100\%}                 & {\color[HTML]{C00000} 29.47} & {\color[HTML]{0070C0} 29.22} & \multicolumn{1}{c|}{\textbf{NODE-ImgNet}}         & \textbf{BRDNet}              & \multicolumn{1}{c|}{\textbf{NODE-ImgNet}}           & \textbf{BRDNet}                \\ \cline{2-8} 
                                    & \textbf{50\%}                  & {\color[HTML]{C00000} 29.46} & {\color[HTML]{0070C0} 29.20} & \multicolumn{1}{c|}{{\color[HTML]{C00000} 0.01}}  & {\color[HTML]{0070C0} -0.02} & \multicolumn{1}{c|}{{\color[HTML]{C00000} -0.03\%}} & {\color[HTML]{0070C0} -0.08\%} \\ \cline{2-8} 
                                    & \textbf{25\%}                  & {\color[HTML]{C00000} 29.35} & {\color[HTML]{0070C0} 28.95} & \multicolumn{1}{c|}{{\color[HTML]{C00000} -0.13}} & {\color[HTML]{0070C0} -0.27} & \multicolumn{1}{c|}{{\color[HTML]{C00000} -0.42\%}} & {\color[HTML]{0070C0} -0.94\%} \\ \cline{2-8} 
                                    & \textbf{10\%}                  & {\color[HTML]{C00000} 29.26} & {\color[HTML]{0070C0} 28.95} & \multicolumn{1}{c|}{{\color[HTML]{C00000} -0.21}} & {\color[HTML]{0070C0} -0.27} & \multicolumn{1}{c|}{{\color[HTML]{C00000} -0.72\%}} & {\color[HTML]{0070C0} -0.94\%} \\ \cline{2-8} 
                                    & \textbf{5\%}                   & {\color[HTML]{C00000} 29.06} & {\color[HTML]{0070C0} 28.90} & \multicolumn{1}{c|}{{\color[HTML]{C00000} -0.41}} & {\color[HTML]{0070C0} -0.32} & \multicolumn{1}{c|}{{\color[HTML]{0070C0} -1.39\%}} & {\color[HTML]{C00000} -1.09\%} \\ \cline{2-8} 
\multirow{-6}{*}{\textbf{Kodak24}}  & \textbf{2\%}                   & {\color[HTML]{C00000} 28.94} & {\color[HTML]{0070C0} 28.50} & \multicolumn{1}{c|}{{\color[HTML]{C00000} -0.53}} & {\color[HTML]{0070C0} -0.72} & \multicolumn{1}{c|}{{\color[HTML]{C00000} -1.81\%}} & {\color[HTML]{0070C0} -2.52\%} \\ \hline
                                    & \textbf{100\%}                 & {\color[HTML]{C00000} 28.29} & {\color[HTML]{0070C0} 28.16} & \multicolumn{1}{c|}{\textbf{NODE-ImgNet}}         & \textbf{BRDNet}              & \multicolumn{1}{c|}{\textbf{NODE-ImgNet}}           & \textbf{BRDNet}                \\ \cline{2-8} 
                                    & \textbf{50\%}                  & {\color[HTML]{C00000} 28.28} & {\color[HTML]{0070C0} 28.10} & \multicolumn{1}{c|}{{\color[HTML]{C00000} -0.01}} & {\color[HTML]{0070C0} -0.06} & \multicolumn{1}{c|}{{\color[HTML]{C00000} -0.04\%}} & {\color[HTML]{0070C0} -0.20\%} \\ \cline{2-8} 
                                    & \textbf{25\%}                  & {\color[HTML]{C00000} 28.22} & {\color[HTML]{0070C0} 27.95} & \multicolumn{1}{c|}{{\color[HTML]{C00000} -0.07}} & {\color[HTML]{0070C0} -0.22} & \multicolumn{1}{c|}{{\color[HTML]{C00000} -0.24\%}} & {\color[HTML]{0070C0} -0.77\%} \\ \cline{2-8} 
                                    & \textbf{10\%}                  & {\color[HTML]{C00000} 28.17} & {\color[HTML]{0070C0} 27.93} & \multicolumn{1}{c|}{{\color[HTML]{C00000} -0.12}} & {\color[HTML]{0070C0} -0.23} & \multicolumn{1}{c|}{{\color[HTML]{C00000} -0.44\%}} & {\color[HTML]{0070C0} -0.83\%} \\ \cline{2-8} 
                                    & \textbf{5\%}                   & {\color[HTML]{C00000} 28.03} & {\color[HTML]{0070C0} 27.90} & \multicolumn{1}{c|}{{\color[HTML]{C00000} -0.26}} & {\color[HTML]{0070C0} -0.26} & \multicolumn{1}{c|}{{\color[HTML]{C00000} -0.91\%}} & {\color[HTML]{0070C0} -0.93\%} \\ \cline{2-8} 
\multirow{-6}{*}{\textbf{CBSD68}}   & \textbf{2\%}                   & {\color[HTML]{C00000} 27.93} & {\color[HTML]{0070C0} 27.58} & \multicolumn{1}{c|}{{\color[HTML]{C00000} -0.36}} & {\color[HTML]{0070C0} -0.59} & \multicolumn{1}{c|}{{\color[HTML]{C00000} -1.30\%}} & {\color[HTML]{0070C0} -2.12\%} \\ \hline
\end{tabular}%
}
\caption{Performance of NODE-ImgNet and BRDNet with decreasing size of training
Dataset. 
Evaluations are shown on CBSD68, Kodak24, and McMaster test sets.}
\label{tab:small_dataset}
\end{table}

\subsection{Ablation Experiments and Demonstration of Model Flexibility}
\label{sec:flexibility}

{
In this subsection, we conducted ablation experiments. The results highlight the efficacy of our approach. Concurrently, these ablation studies also reveal a significant advantage of our NODE structure: It offers flexibility in choosing the value for the hyper-parameter of time integration steps N, which facilitating a balance between training time and accuracy.
}


{
The novelty of our model stems from the combination of the NODE model with the CNN network. This necessitates us to consider two aspects when designing ablation experiments: firstly, by removing the NODE structure to demonstrate the denoising capability of the original Vector field CNN structure; and secondly, by adjusting the time step N in NODE to showcase the model's denoising ability under different N scenarios. For the former approach, we design two models. In the first model, we entirely remove the NODE structure, retaining only the vector field CNN framework. We refer to this model as the `Vector field (N=0)'. In the second model, while removing the NODE structure, we also introduce a residual block to bridge the input and output, thereby enhancing the original CNN network. We designate this model as the `Vector field (N=1)'. The rationale behind this design is that this residual structure resembles the NODE structure when N=1. Theoretically, NODE at N=1 is equivalent to not partitioning time, and thus, we disregard this scenario for NODE-ImgNet. For the latter approach, we test the denoising capability of NODE-ImgNet by solely adjusting the time step N from 2 to 8, examining its performance under various time steps.}

{For a fair comparison, all of the aforementioned models are trained based on the full Waterloo dataset with a fixed noise level of $\sigma=50$. The final experimental results are shown in \Cref{N PSNR pic}. On the x-axis, the labels N=0 and N=1 correspond to the Vector field (N=0) and Vector field (N=1) models respectively, while labels N=2 to N=8 represent the NODE-ImgNet with time steps from N=2 to N=8. In the figure, the model's performance is depicted by a blue curve, while the training time is represented by a red curve.

Comparing the results of Vector field (N=0, 1) with NODE-ImgNet N=2-8 underscores the efficacy of incorporating NODE with an appropriate number of steps, thus validating the effectiveness of our structure. Further, when observing the results of NODE-ImgNet from N=2 to N=8, it's evident that both the training time (for 100 batches) and the PSNR value exhibit an almost linear relationship with the value of N. This demonstrates the flexibility of our model, allowing users to make a trade-off between performance and training time by simply adjusting the value of N.

However, it's crucial to point out that as the value of N increases, the growth rate for PSNR is relatively more gradual than that for the training time. Nonetheless, it's noteworthy that with N=2, NODE-ImgNet already surpasses BRDNet under an identical noise setting.
}
}


\begin{figure}[hbt!]
    \centering
    \begin{minipage}{0.46\textwidth}
        \includegraphics[width=\textwidth]{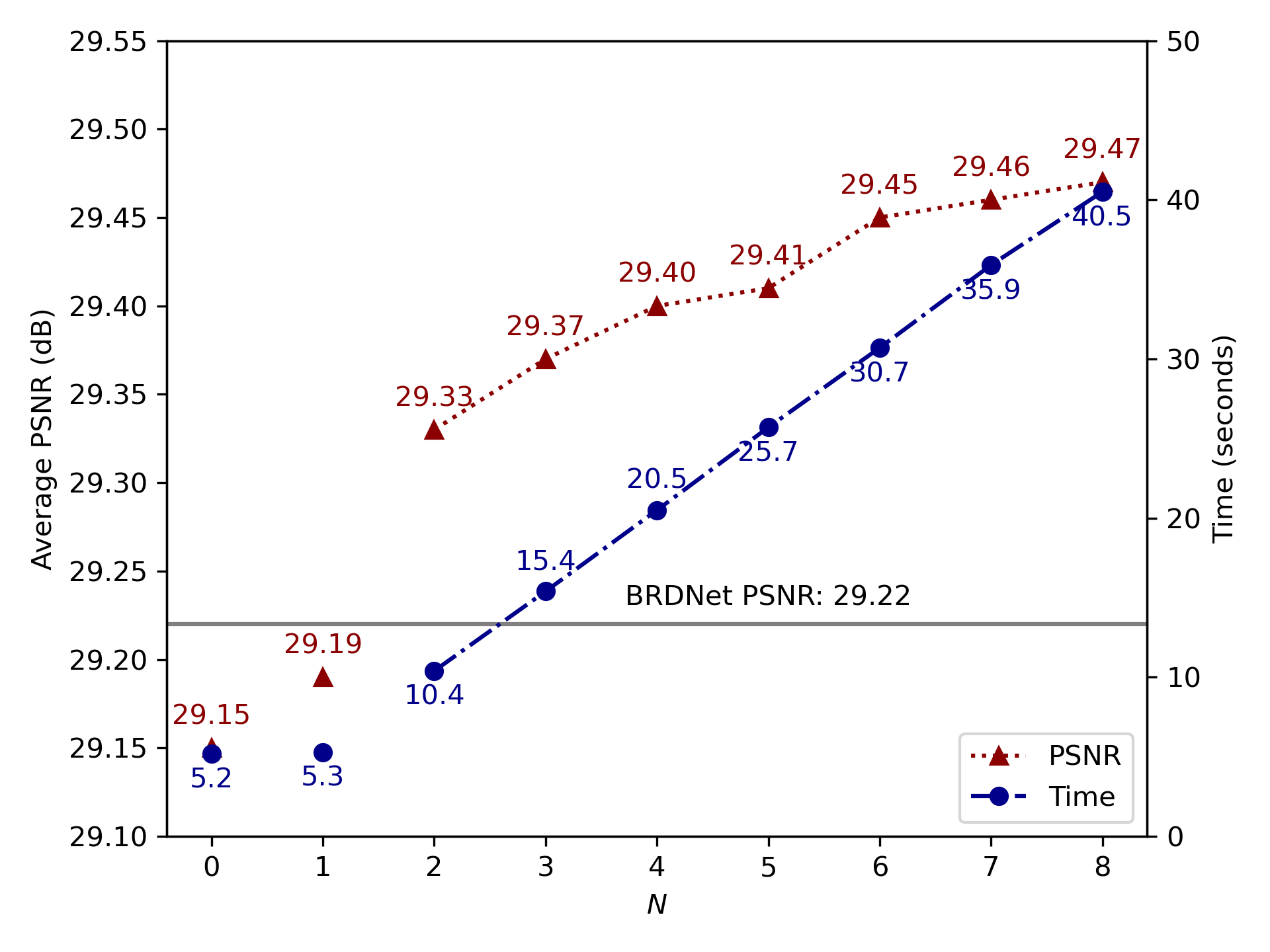}
        \caption{Average PSNR for Kodak24 test set and training time for $100$ batches under varying $N$ values.}
        \label{N PSNR pic}
    \end{minipage}
    \hspace{1em} 
    \begin{minipage}{0.46\textwidth}
        \includegraphics[width=\textwidth]{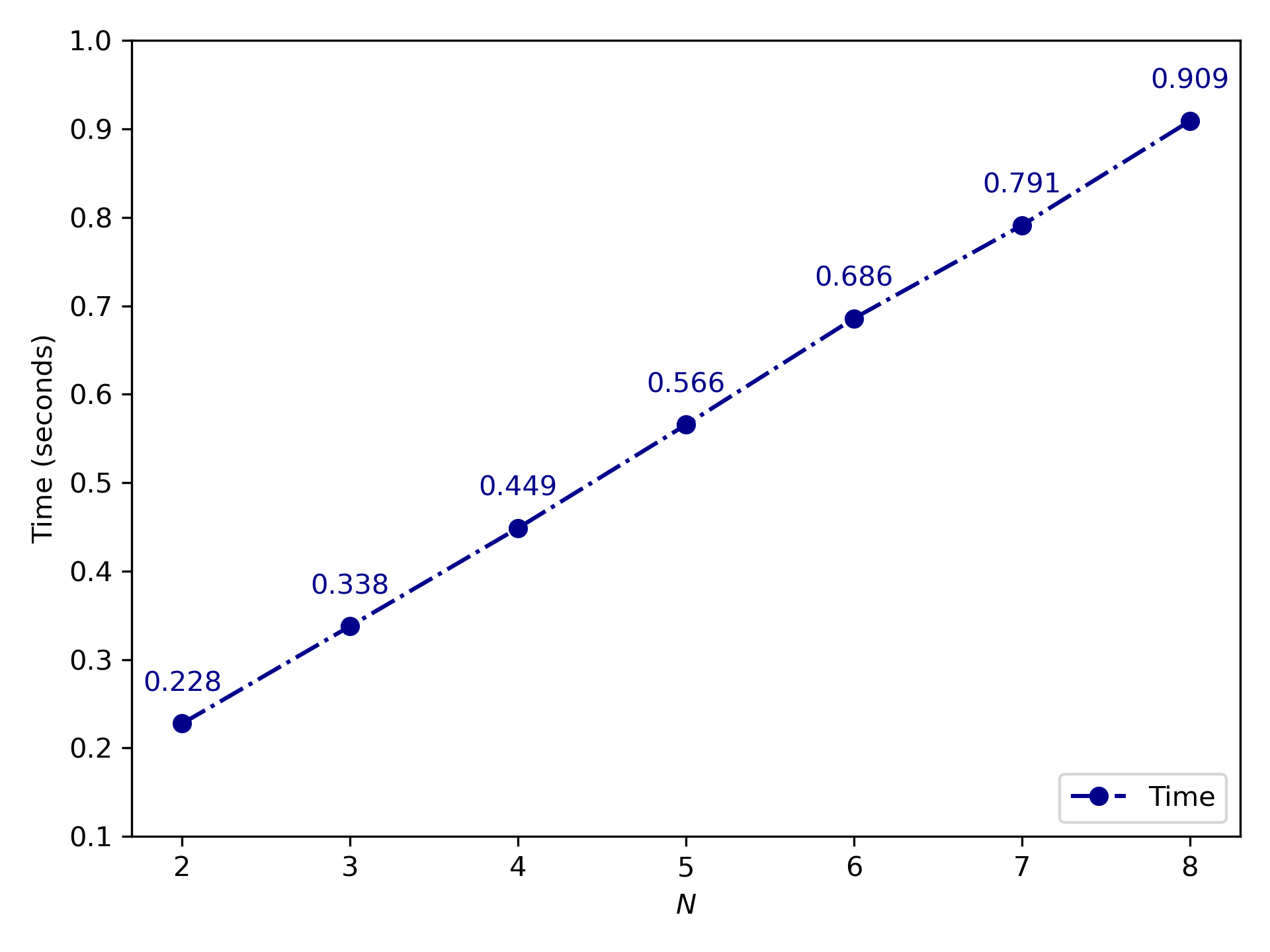}
        \caption{Test time versus $N$ values on a $1024 \times 1024$ color image using NODE-ImgNet.}
        \label{fig: test time}
    \end{minipage}
\end{figure}

\subsection{Computational Costs}
\label{sec:computational costs}
Testing time and the number of parameters are also important metrics for evaluating models \cite{schmidt2014shrinkage}. 
In this subsection, we compare our model NODE-ImgNet with various other baseline models including FFDnet, DnCNN, MWDCNN, and BRDNet in terms of evaluation/testing time and the number of parameters.
The testing times are obtained on testing color images of size $256\times256$, $512\times512$, and $1024\times1024$ with Gaussian fixed noise level $\sigma = 50$. To ensure fair comparison, we rebuild these models and test them on the same device mentioned in section \ref{sec:Expaermental setting}. The results are shown in \Cref{tab:test-time}.  We also plot the testing time for denoising the same $1024\times1024$ color image versus
$N$,  as shown in Figure \ref{fig: test time}. 
{
We observe that when $N=2$, the testing time of NODE-ImgNet is comparable with BRDNet whereas NODE-ImgNet performs better and has less number of parameters.} 


\begin{table}[]
\renewcommand{\arraystretch}{1.5} 
\resizebox{\columnwidth}{!}{%
\begin{tabular}{|cc|c|c|c|c|c|ccc|}
\hline
\multicolumn{2}{|c|}{\multirow{2}{*}{\textbf{Methods}}}                     & \multirow{2}{*}{\textbf{DnCNN}} & \multirow{2}{*}{\textbf{ADNet}} & \multirow{2}{*}{\textbf{DudeNet}} & \multirow{2}{*}{\textbf{MWDCNN}} & \multirow{2}{*}{\textbf{BRDNet}} & \multicolumn{3}{c|}{\textbf{NODE-ImgNet}}                                                                                     \\ \cline{8-10} 
\multicolumn{2}{|c|}{}                                                      &                                 &                                 &                                   &                                  &                                  & \multicolumn{1}{c|}{\textbf{N=2, 128 channels}} & \multicolumn{1}{c|}{\textbf{N=8, 128 channels}} & \textbf{N=8, 64 channels} \\ \hline
\multicolumn{2}{|c|}{\textbf{Parameters}}                                   & 0.56M                           & 0.52 M                          & 1.03M                             & 0.50M                            & 1.11M                            & \multicolumn{1}{c|}{1.04M}                      & \multicolumn{1}{c|}{1.04M}                      & 0.26M                     \\ \hline
\multicolumn{1}{|c|}{\multirow{3}{*}{\textbf{size}}} & \textbf{256 × 256}   & 0.007s                          & 0.012s                          & 0.014                             & 0.027s                           & 0.012s                           & \multicolumn{1}{c|}{0.017s}                     & \multicolumn{1}{c|}{0.072}                      & 0.052s                    \\ \cline{2-10} 
\multicolumn{1}{|c|}{}                               & \textbf{512 × 512}   & 0.024s                          & 0.032s                          & 0.047s                            & 0.147s                           & 0.047s                           & \multicolumn{1}{c|}{0.060s}                     & \multicolumn{1}{c|}{0.254s}                     & 0.148s                    \\ \cline{2-10} 
\multicolumn{1}{|c|}{}                               & \textbf{1024 × 1024} & 0.103s                          & 0.142s                          & 0.221s                            & 0.877s                           & 0.197s                           & \multicolumn{1}{c|}{0.228s}                     & \multicolumn{1}{c|}{0.887s}                     & 0.4763s                   \\ \hline
\end{tabular}%
}
\caption{Comparison of testing time for various models using different input image sizes.}
\label{tab:test-time}
\end{table}

It is worth mentioning that BRDNet utilizes a channel number of $64$ in its intermediate layers. To provide a comparison point, we also train NODE-ImgNet models with the same intermediate channel number of $64$ and $N=8$.

In this scenario, the vector field network employed by NODE-ImgNet is a proper subset of BRDNet, resulting in NODE-ImgNet utilizing only $23\%$ of
the parameters found in BRDNet.
The performance of NODE-ImgNet, with a channel number of $64$, is presented in \Cref{NODE-ImgNet-c64}. We also include a comparison to other baseline models for three color image test datasets.
Upon examining the results, our model demonstrates superior performance compared to all the baseline models for the chosen high levels of noise.

The results suggest that NODE-ImgNet exhibits significant potential for image denoising on portable devices like smartphones, drones, embedded systems, and cameras.

\begin{table}[h!]
\renewcommand{\arraystretch}{1.5} 

\resizebox{\columnwidth}{!}{%
\begin{tabular}{|c|c|c|c|c|c|c|c|c|}
\hline
\textbf{Datasets}                   & {\color[HTML]{2E2E2E} \textbf{Noise Level}}         & \textbf{DnCNN \cite{zhang2017beyond}} & \textbf{FFDNet \cite{zhang2018ffdnet}} & \textbf{GradNet \cite{liu2020gradnet}} & \textbf{ADNet \cite{tian2020attention}} & \textbf{DudeNet \cite{tian2021designing}} & \textbf{BRDNet \cite{tian2020image}} & \textbf{NODE-ImgNet-S}       \\ \hline
                                    & {\color[HTML]{2E2E2E} \textbf{$\sigma   = 50$}} & 28.01                                 & 27.96                                  & 28.12                                  & 28.04                                   & 28.09                                     & {\color[HTML]{00B0F0} 28.16}         & {\color[HTML]{FF0000} 28.19} \\ \cline{2-9} 
\multirow{-2}{*}{\textbf{CBSD68}}   & {\color[HTML]{2E2E2E} \textbf{$\sigma = 75$}}   & -                                     & 26.24                                  & -                                      & 26.33                                   & 26.40                                     & {\color[HTML]{00B0F0} 26.43}         & {\color[HTML]{FF0000} 26.54} \\ \hline
                                    & {\color[HTML]{2E2E2E} \textbf{$\sigma   = 50$}} & {\color[HTML]{2E2E2E} 29.02}          & {\color[HTML]{2E2E2E} 28.99}           & {\color[HTML]{00B0F0} 29.23}           & 29.1                                    & 29.1                                      & 29.22                                & {\color[HTML]{FF0000} 29.30} \\ \cline{2-9} 
\multirow{-2}{*}{\textbf{Kodak24}}  & {\color[HTML]{2E2E2E} \textbf{$\sigma = 75$}}   & {\color[HTML]{2E2E2E} -}              & {\color[HTML]{2E2E2E} 27.25}           & {\color[HTML]{2E2E2E} -}               & 27.4                                    & 27.39                                     & {\color[HTML]{00B0F0} 27.49}         & {\color[HTML]{FF0000} 27.67} \\ \hline
                                    & {\color[HTML]{2E2E2E} \textbf{$\sigma   = 50$}} & {\color[HTML]{2E2E2E} 29.21}          & {\color[HTML]{2E2E2E} 29.21}           & {\color[HTML]{2E2E2E} 29.39}           & 29.36                                   & -                                         & {\color[HTML]{00B0F0} 29.52}         & {\color[HTML]{FF0000} 29.57} \\ \cline{2-9} 
\multirow{-2}{*}{\textbf{McMaster}} & {\color[HTML]{2E2E2E} \textbf{$\sigma = 75$}}   & {\color[HTML]{2E2E2E} -}              & {\color[HTML]{2E2E2E} 27.29}           & {\color[HTML]{2E2E2E} -}               & 27.53                                   & -                                         & {\color[HTML]{00B0F0} 27.72}         & {\color[HTML]{FF0000} 27.81} \\ \hline
\end{tabular}%
}
\caption{Comparison of NODE-ImgNet (channel number $64$) and various other models for Gaussian color image denoising at two high noise levels.}
\label{NODE-ImgNet-c64}
\end{table}



\section{Conclusion}\label{sec:discussion}

Inspired by the traditional PDE approach to image denoising, we propose an image denoising model referred as NODE-ImgNet in this paper.  The structure of NODE-ImgNet employed in this paper combines NODEs
with CNN blocks and achieves enhanced accuracy and parameter efficiency. 
The experimental data suggest that NODE-ImgNet consistently performs at a high level across diverse image denoising scenarios. Additionally, this model exhibits proficiency in learning from limited training datasets and provides flexibility in model complexity tuning to meet specific task parameters. {Nevertheless, the present framework does not integrate important image metrics, e.g., gradient, divergence, etc. This omission may constrains the model's full potential in optimizing denoising outcomes. It is also noteworthy that augmenting the model with pre-processed image data using traditional models could further enhance its performance capabilities.}

{
One of our future objectives is to enhance the efficiency of our model by explicitly incorporating spatial derivatives into the vector field. Specifically, we aim to expand the application of our proposed architecture to tackle increasingly complex real-world image denoising tasks, including denoising low-light conditioned and blurred images. Moreover, we plan to investigate the integration of advanced techniques such as attention mechanisms \cite{bahdanau2014neural} and U-Net \cite{ronneberger2015u} within the vector field, further augmenting the effectiveness of our model.}

{
Adopting our proposed framework offers several benefits to users and other models: 1) Proper integration of other benchmark models to the vector field of NODE-ImgNet has the potential to significantly enhance the performance of existing benchmark structures; 2) Its inherent versatility allows for a wide range of applications, including image segmentation and high-resolution reconstruction.}


\section*{Acknowledgements}
We would like to express our gratitude to the Oxford Mathematics Institute for providing server support and technical assistance. We are also grateful for Baoren Xiao (UCL)'s assistance with some of the numerical experiments. 

\section*{Funding}
HN is supported by the Engineering and Physical Sciences Research Council (EPSRC) [grant number EP/S026347/1] and the Alan Turing Institute under the EPSRC grant [grant number EP/N510129/1].

\newpage
\appendix






\pagebreak
\bibliographystyle{elsarticle-num}





\end{document}